%% file: main.tex
\documentclass{article}
\usepackage{iclr2027_provisional,times}
\usepackage{hyperref}
\usepackage{url}
\usepackage{booktabs}
\usepackage{multirow}
\usepackage{graphicx}
\usepackage{amsmath,amssymb}
\usepackage{xspace}
\usepackage{xcolor}
\usepackage{microtype}
\usepackage{enumitem}
\usepackage{float}
\usepackage{array}
\usepackage{tabularx}

\newcommand{\benchname}{\textsc{MaliciousSkillBench}\xspace}

\hypersetup{
  colorlinks=true,
  linkcolor=blue,
  citecolor=blue,
  urlcolor=blue,
  pdftitle={MaliciousSkillBench: A Comprehensive Benchmark for Malicious Agent Skill Detection},
  pdfauthor={Yue Wang, Yi Liu, Gelei Deng, Ying Zhang, Yuekang Li, Zhenyu Chen, Leo Zhang},
  pdfsubject={MaliciousSkillBench arXiv preprint}
}

\title{MaliciousSkillBench:\\ A Comprehensive Benchmark for\\ Malicious Agent Skill Detection}

\author{Yue Wang$^{1}$\hspace{0.75em} Yi Liu$^{2}$\hspace{0.75em} Gelei Deng$^{3}$\hspace{0.75em} Ying Zhang$^{4}$\hspace{0.75em} Yuekang Li$^{5}$\hspace{0.75em} Zhenyu Chen$^{1}$\hspace{0.75em} Leo Zhang$^{2}$}

\iclrpreprintaffiliations{%
$^{1}$State Key Laboratory for Novel Software Technology, Nanjing University\\[-0.12em]
$^{2}$Griffith University \quad
$^{3}$Nanyang Technological University \quad
$^{4}$Wake Forest University\\[-0.12em]
$^{5}$University of New South Wales}

\iclrpreprintcopy

\begin{document}
\maketitle

\input{sections/00_abstract}
\input{sections/01_introduction}
\input{sections/02_related_work}
\input{sections/03_why_comprehensive}
\input{sections/04_building}
\input{sections/05_threat_landscape}
\input{sections/06_detection}
\input{sections/07_discussion}
\input{sections/08_limitations}
\input{sections/09_conclusion}
\input{sections/10_statements}

\input{references_manual}

\appendix
\input{appendix/appendix}

\end{document}

%% file: sections/00_abstract.tex
\begin{abstract}
Agent Skills extend LLM agents with reusable instruction packages that may also include scripts, resources, and service configuration. This creates a direct distribution channel for malicious behavior, yet existing malicious-Skill datasets are fragmented across sources, artifact formats, evidence regimes, and benign coverage; duplicated and structurally related content further complicates direct aggregation and evaluation. We present \benchname, a comprehensive benchmark for malicious Agent Skill detection. We consolidate 13 public sources, 11 of which contribute Core malicious artifacts, and reduce 8,414 raw malicious records to 7,539 normalized-unique identities in 4,588 operational structural families. After conservative cross-label conflict exclusion, the primary benchmark contains \textbf{9,740 Skills: 7,505 malicious and 2,235 benign}. To characterize its coverage, we harmonize \textbf{11 attack categories} for 4,983 malicious identities with supported source-native mappings and find substantial differences in threat composition across sources. We then evaluate three learned text detectors and three off-the-shelf Skill scanners. Learned detectors achieve 0.882--0.932 Random Macro-F1 but only 0.653--0.665 under Source-Disjoint evaluation; the strongest word TF--IDF SVM scores \textbf{0.932/0.916/0.665} on Random/structural-disjoint/Source-Disjoint while retaining 95.6\% malicious recall but producing 62.4\% benign FPR on held-out sources. Off-the-shelf scanners occupy different but also unsatisfactory operating regimes, reducing false positives only at the cost of sharply lower malicious recall. Together, these results show that reliable malicious-Skill detection requires both broader cross-source benchmark coverage and evaluation that jointly measures attack detection and benign over-flagging.
\textbf{Resources:} \href{https://protectskills.github.io/MaliciousSkillBench/}{Project Page}.
\end{abstract}

%% file: sections/01_introduction.tex
\section{Introduction}
\label{sec:intro}

Large language model (LLM) agents increasingly acquire reusable capabilities through \emph{Skills}: installable packages that combine natural-language instructions with scripts, templates, resources, and service configuration. A malicious Skill can therefore act as trusted procedural authority inside the agent's workflow. Recent studies show that malicious Skills can steal credentials, manipulate agent behavior, introduce triggered backdoors, or conceal unsafe side effects \citep{guo2026malskillbench,zhuang2026agenttrap,liu2026maliciouswild,feng2026skilltrojan}. This makes pre-installation malicious-Skill detection an increasingly important security problem.

The available datasets do not yet provide a single, reliable basis for studying that problem. Existing Skill-security resources were created for different purposes and release different objects: complete Skill artifacts, constructed attack variants, runtime-verified cases, vulnerability records, marketplace observations, or scanner-derived signals \citep{guo2026malskillbench,zhuang2026agenttrap,liu2026maliciouswild,ning2026skillharm,jin2026skillsafety}. Their labels are supported by different mixtures of construction, human review, runtime verification, static analysis, and automated scanning, and their benign coverage is uneven. Sources also overlap: identical or formatting-equivalent Skills recur across datasets, related variants reuse common scaffolds, and the same normalized content can receive conflicting labels. Simply concatenating published rows therefore overstates independent coverage and can produce misleading evaluation splits.

We build \benchname to turn this fragmented landscape into a common detection benchmark. Here, \emph{comprehensive} refers to broad, traceable cross-source consolidation of Skill artifacts from 13 frozen public sources. We freeze 13 public sources and retain 11 that satisfy a conservative Core-malicious artifact rule. We canonicalize heterogeneous records while preserving source and evidence semantics, deduplicate exact and formatting-equivalent content, model broader structural reuse separately from attack semantics, and exclude cross-label conflicts before evaluation. The resulting benchmark contains \textbf{9,740 normalized-unique Skills: 7,505 malicious and 2,235 benign}. Its malicious side originates from 8,414 raw Core artifacts, which reduce to 7,562 exact-unique and 7,539 normalized-unique identities organized into 4,588 operational structural families.

\begin{figure}[t]
    \centering
    \includegraphics[width=0.93\linewidth]{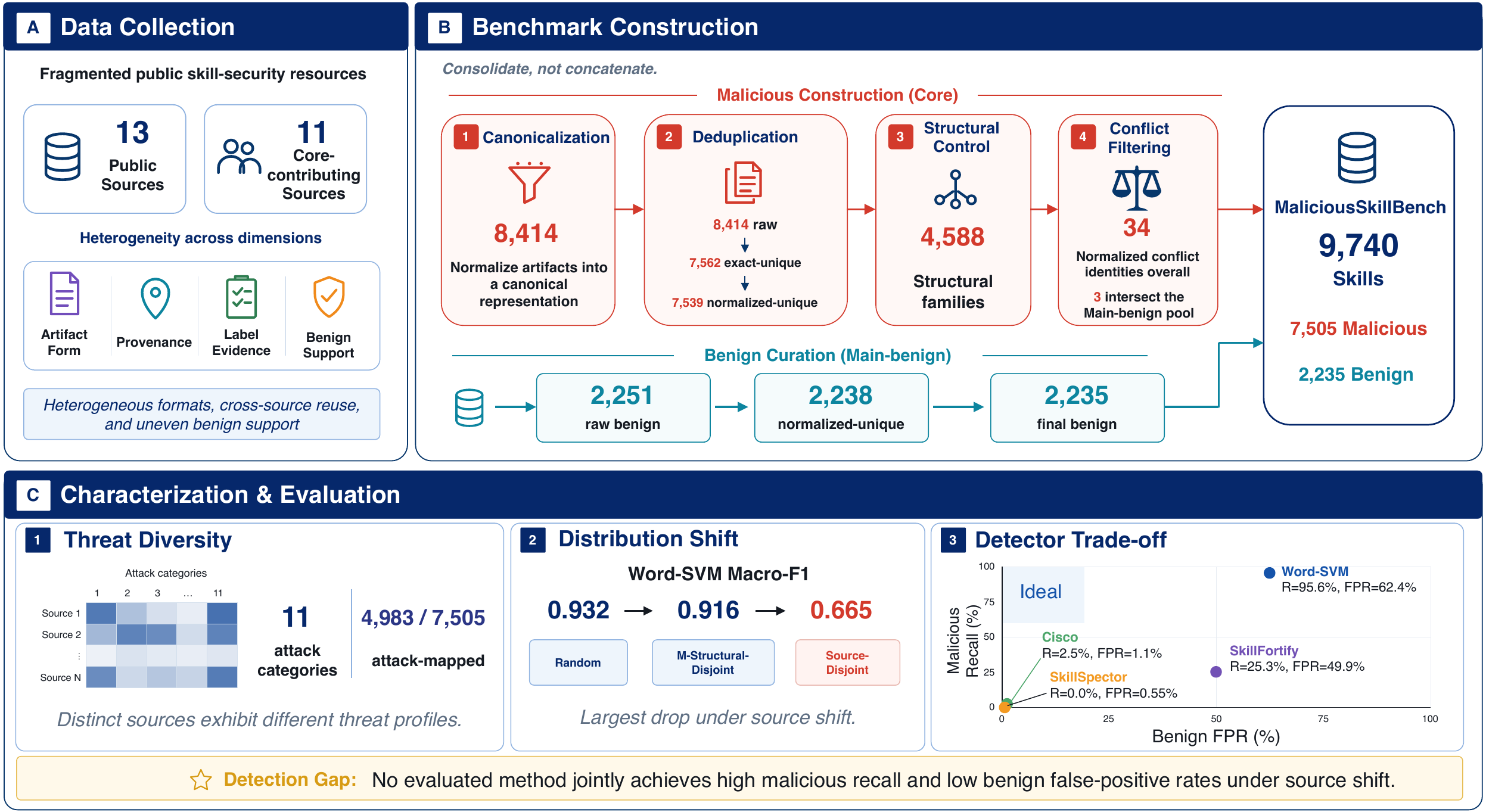}
    \caption{Overview of \benchname. \textbf{(A)} Fragmented sources differ in artifacts, provenance, evidence, and benign support. \textbf{(B)} Controlled canonicalization, deduplication, structural grouping, benign curation, and conflict filtering yield 9,740 normalized-unique Skills (7,505 malicious; 2,235 benign). \textbf{(C)} Characterization and held-out-source evaluation expose heterogeneous attack coverage and distinct detector recall/FPR regimes; the source--attack matrix is schematic (quantitative results in Figure~\ref{fig:threat_landscape}).}
    \label{fig:overview}
\end{figure}

The consolidated benchmark also makes it possible to ask what malicious behavior existing resources actually cover. Using only documented source-native labels and deterministic strong-semantic mappings, we harmonize \textbf{11 attack categories for 4,983 of the 7,505 malicious identities}. Their distributions differ sharply across sources, so individual datasets emphasize different slices of the threat landscape.

Finally, \benchname changes the conclusion one would draw about detector performance. Across three learned static text baselines, Random Macro-F1 is \textbf{0.882--0.932}, but Source-Disjoint Macro-F1 falls to \textbf{0.653--0.665}. The strongest word TF--IDF SVM scores \textbf{0.932/0.916/0.665} under Random, malicious-structural-disjoint, and Source-Disjoint evaluation. On held-out sources it still detects 95.6\% of malicious Skills, yet flags 62.4\% of benign Skills as malicious: malicious recall remains high while benign over-flagging dominates. Three off-the-shelf Skill-security scanners reveal the opposite failure regime: configurations with very low benign false-positive rates detect only a small fraction of malicious Skills, while the more sensitive scanner incurs substantial benign false positives. No evaluated detector simultaneously achieves high malicious recall and low benign FPR across held-out sources.

\textbf{Our contributions are:}
\begin{itemize}[leftmargin=*,nosep]
    \item \textbf{A comprehensive malicious-Skill detection benchmark.} We consolidate 13 public sources into a traceable artifact-level resource, preserving heterogeneous evidence while controlling exact/normalized duplication, structural reuse, and cross-label conflicts; the frozen primary benchmark contains 9,740 Skills.
    \item \textbf{A source-aware threat characterization.} We harmonize 11 attack categories for the supported mapped subset and quantify source, provenance, attack, and derived-impact coverage, revealing substantial heterogeneity across existing resources.
    \item \textbf{A unified evaluation of learned and off-the-shelf detectors.} Random, structural-disjoint, and source-disjoint tests reveal substantial cross-source degradation despite strong random-split scores: learned models over-flag unfamiliar benign Skills, while current scanners reduce false positives largely by sacrificing malicious recall.
\end{itemize}

%% file: sections/02_related_work.tex
\section{Background and Related Work}
\label{sec:related}

\subsection{Agent and Tool Security}
Tool-using agents inherit security failures from untrusted observations, tool interfaces, and persistent state. InjecAgent and AgentDojo established indirect prompt injection in realistic tool-mediated workflows, while Agent Security Bench broadened evaluation to attacks on prompts, tools, and memory \citep{zhan2024injecagent,debenedetti2024agentdojo,zhang2024asb}. More recent MCP studies show that even tool metadata can become an attack channel \citep{wang2025mcptox}. Agent Skills create a related but distinct trust boundary: the Skill itself is installed as reusable procedural authority and may combine instructions, helper code, resources, and permissions. This creates a distinct security question about the trustworthiness of an installed capability package before it becomes reusable procedural authority.

\subsection{Skill-Specific Threats, Datasets, and Defenses}
Recent work studies complementary parts of the Skill attack surface. Skill-Inject evaluates prompt injections carried by Skill files, SkillTrojan studies triggered backdoors, and SkillSafetyBench and SkillHarm study Skill-mediated unsafe behavior and harmful capability use \citep{schmotz2026skillinject,feng2026skilltrojan,jin2026skillsafety,ning2026skillharm}. Large-scale empirical studies also characterize marketplace-scale Skill vulnerabilities and credential leakage \citep{liu2026skillswildvuln,chen2026credentialleak}. Artifact-oriented datasets cover other parts of the space: \citet{liu2026maliciouswild} behaviorally verifies marketplace samples; AgentTrap evaluates complete runtime trajectories; and MalSkillBench combines generated runtime-verified attacks, wild samples, and matched benign Skills \citep{zhuang2026agenttrap,guo2026malskillbench}. Scanner-oriented resources use still different supervision: ClawHub Security Signals releases a large silver-standard marketplace snapshot, SkillTrustBench separates normal, suspicious, and malicious cases for scanner evaluation, and SkillFortifyBench provides a deterministic multi-format synthetic testbed \citep{koc2026clawhubsignals,skilltrustbench2026,bhardwaj2026skillfortify}.

Defenses are similarly heterogeneous. RouteGuard uses internal model signals for pre-execution poisoning detection, Locate-and-Judge uses attention to scale marketplace triage, SkillGate combines lexical prefiltering with LLM judgment, and Runtime Skill Audit and SkillDetonate move toward behavior-centric execution-time evidence \citep{xiao2026routeguard,etteib2026locatejudge,yang2026skillgate,lan2026runtimeaudit,ji2026cloakdetonate}. Other work exposes blind spots beyond plain text, including compositional risk among individually safe Skills and malicious instructions hidden in visual resources \citep{wang2026skillsreact,jia2026seeing}. ColluSkill further studies adversarial cross-Skill composition and reports that harmful workflows can evade scanners that inspect Skills individually \citep{zeng2026colluskill}. \benchname is complementary to these attack and defense proposals by consolidating heterogeneous public resources into a common artifact-level detection benchmark and evaluating both learned and off-the-shelf detectors under the same frozen data contract.

\subsection{Benchmark Artifacts and Distribution Shift}
A broader benchmark literature shows why this consolidation must control both duplication and domain-specific shortcuts. Annotation artifacts can make labels predictable from unintended cues \citep{gururangan2018annotation}; WILDS demonstrates degradation under naturally occurring distribution shifts, and DomainBed shows that domain-generalization conclusions depend strongly on standardized evaluation protocols \citep{koh2021wilds,gulrajani2021domainbed}. We use source identity and construction provenance for audit and stratification and exclude them from detector features. Our held-out-source evaluation measures source-conditioned stress under coupled source factors; normalized deduplication, malicious-structural-disjoint evaluation, class balancing, and scaffold sanitization probe distinct shortcut channels.

%% file: sections/03_why_comprehensive.tex
\section{Why a Comprehensive Benchmark?}
\label{sec:why_comprehensive}

Existing malicious-Skill datasets were built under different assumptions and cannot be treated as interchangeable rows. Our 13-source registry exposes four concrete deficiencies that a detection benchmark must address.

\paragraph{Existing coverage is fragmented.}
Prior work studies complementary slices of the Skill threat surface: generated and runtime-verified malicious artifacts, in-the-wild marketplace samples, triggered backdoors, harmful capability use, vulnerability records, and scanner-oriented resources \citep{guo2026malskillbench,zhuang2026agenttrap,liu2026maliciouswild,feng2026skilltrojan,ning2026skillharm,jin2026skillsafety}. Their released units and labels are not equivalent. A complete Skill artifact supported by runtime evidence is a different observation from a vulnerability record or an automated scanner verdict. A comprehensive benchmark must reconcile these resources without erasing source, artifact, provenance, or evidence semantics.

\paragraph{More rows do not necessarily mean more independent coverage.}
In the frozen collection, 8,414 Core malicious records reduce to 7,562 exact-unique and 7,539 normalized-unique Skill contents. Broader static similarity further partitions the malicious side into 4,588 operational structural families, while 34 normalized identities collide across malicious and benign labels. These are different phenomena---content identity, scaffold reuse, and label inconsistency---and they require different controls. Simply concatenating all rows would inflate apparent scale and can place reused content on both sides of a split.

\paragraph{Detection also requires credible benign data.}
The source registry is strongly asymmetric: several resources contain malicious Skills only, and benign labels vary in strength. After conservative eligibility and conflict exclusion, the primary benchmark contains 7,505 malicious and 2,235 Main-benign normalized-unique Skills. This matters operationally because source, construction procedure, and class composition are coupled. A detector can achieve high apparent attack coverage simply by broadly flagging unfamiliar artifacts. Evaluating deployable detection therefore requires high-confidence benign examples from multiple source conventions.

\paragraph{Random splits are an incomplete test of deployment robustness.}
Random splitting is useful as a reference, but it can distribute source conventions and structurally related variants across train and test. Cross-domain benchmark practice similarly distinguishes in-distribution accuracy from performance under domain shift \citep{koh2021wilds,gulrajani2021domainbed}. For malicious Skills, source identity bundles collection venue, attack construction, documentation style, labeling policy, provenance, and benign population. We therefore need complementary views of difficulty: random evaluation, malicious structural-family disjointness, and held-out-source evaluation. The last measures source-conditioned stress under coupled source factors.

These deficiencies translate directly into the design of \benchname: a common artifact-level unit, evidence-preserving canonicalization, explicit identity/reuse/conflict control, a high-confidence benign pool, and evaluation protocols that distinguish conventional random performance from structural and source-conditioned generalization. Section~\ref{sec:benchmark} describes the construction pipeline.

%% file: sections/04_building.tex
\section{Building \benchname}
\label{sec:benchmark}

We construct \benchname with four consolidation stages (Figure~\ref{fig:overview}): qualify source artifacts, map records into an evidence-preserving schema, control identity and structural reuse, and finalize the detection set after conflict exclusion. Detailed controlled vocabularies, lineage records, and release rules are deferred to the appendix.

\subsection{Source Collection and Artifact Eligibility}
\label{sec:benchmark_sources}

\paragraph{Collect broadly, qualify conservatively.}
We freeze 13 public malicious-Skill datasets and security corpora and record source revision, license, redistribution status, reported/acquired counts, and artifact availability. Across all benchmark roles, the canonical manifest contains 182,699 records. A source contributes to the Core malicious benchmark only when we can recover an actual Skill artifact, preserve the source-native malicious claim and its supporting evidence, keep the artifact inert as detector input, and trace it to a frozen source revision. Security-relevant task pairs, vulnerability-only rows, scanner verdicts, environment fixtures, and non-Skill configurations are retained as Auxiliary under their original source semantics.

Eleven of the 13 sources satisfy this Core rule (Table~\ref{tab:core_sources}), contributing \textbf{8,414 raw malicious Skill artifacts}. Their provenance spans wild, test-fixture, injected/backdoored, and synthetic collections, and their benign support is highly uneven. This heterogeneity is retained so later analyses can distinguish what was observed, constructed, or inferred.

\input{tables/core_sources}

\paragraph{Recover missing artifacts without relaxing eligibility.}
Two sources require a documented acquisition exception. The public releases of SRC002 (\emph{MaliciousAgentSkillsBench}) and SRC004 (\emph{SkillLeakBench}) \citep{chen2026credentialleak} provide study metadata/labels but not the raw malicious Skill artifacts needed for our static benchmark. We keep those public revisions as the source of record and link eligible rows to author-provided historical research snapshots, recovering \textbf{157 SRC002} and \textbf{79 SRC004} malicious artifacts under the same Core rule. Four unresolved SRC004 multi-Skill cases and its vulnerability-only records remain outside Core. Benchmark membership and redistribution permission are tracked separately.

\subsection{Canonicalization without Erasing Evidence}
\label{sec:benchmark_canonical}

Every acquired record receives a stable canonical ID while retaining source ID, source label, source revision, artifact type, provenance, label strength, evidence type, hashes, and available lineage metadata. Source-specific labels are mapped conservatively: a record supported only as \emph{vulnerable}, \emph{suspicious}, dual-use, or scanner-flagged is not upgraded to malicious ground truth, and unresolved provenance or parentage remains unresolved. The complete vocabulary records intent, provenance, and evidence as distinct fields (Appendix~\ref{app:taxonomy}).

This common schema prevents source-native row count, variant count, and independent content identity from being conflated. Published base/variant links are retained when explicit; similarity is used only for structural reuse analysis and never establishes causal lineage. Throughout acquisition and canonicalization, potentially malicious artifacts are handled as inert static data: the pipeline copies or parses text and metadata but does not execute Skill code, helpers, payloads, URLs, installers, or embedded instructions.

\subsection{Deduplication and Structural Reuse Control}
\label{sec:benchmark_dedup}

We separate three forms of relatedness because they answer different benchmark questions. \emph{Exact identity} is SHA-256 equality over acquired Skill content. \emph{Normalized identity} applies deterministic conservative text normalization to collapse formatting-equivalent content while preserving substantive text. These stages reduce \textbf{8,414 raw Core malicious artifacts} to \textbf{7,562 exact-unique} and \textbf{7,539 normalized-unique} malicious identities. Per-source normalized counts in Table~\ref{tab:core_sources} can overlap because the same identity may occur in multiple sources.

Broader scaffold reuse is handled separately from identity and threat semantics. On one representative per normalized malicious identity, a frozen static-similarity pipeline at threshold 0.68 yields \textbf{4,588 operational structural families}. We use these families only for reuse audits and malicious-side structural-disjoint splitting; they encode operational structural reuse with no attack, campaign, actor, or threat-class semantics. A blinded positive-only review of 72 sampled within-family pairs provides a bounded coherence check, with the full method and limitations reported in Appendix~\ref{app:reuse}.

\subsection{Benign Pool, Label Conflicts, and Final Benchmark}
\label{sec:benchmark_benign}

A detection benchmark also needs credible negatives. We separate \textbf{48,217 benign candidates} into Main and Auxiliary pools according to artifact availability and source evidence. Main benign records require an actual Skill artifact plus strong or moderate benign evidence; marketplace-unflagged, scanner-clean, silver, or otherwise weak negatives remain Auxiliary. This yields 2,251 raw Main-benign artifacts (2,238 normalized-unique) and 45,966 raw Auxiliary-benign artifacts (44,327 normalized-unique). Auxiliary examples remain useful for audit and future work but are not used as primary benign ground truth.

Finally, we compare exact and normalized identities across malicious and benign pools. \textbf{Thirty-four normalized identities} receive conflicting malicious/benign labels across frozen sources. We exclude every conflicting identity from primary evaluation and retain it for consistency auditing, avoiding new automatic adjudication. The resulting frozen benchmark contains \textbf{9,740 normalized-unique Skills: 7,505 malicious and 2,235 Main benign}. Every evaluation protocol starts from this same master table; no split-specific relabeling or synthetic oversampling changes the underlying benchmark population.

%% file: tables/core_sources.tex
\begin{table}[H]
\centering
\caption{Coverage of the 11 Core-contributing sources in the frozen benchmark. ``Norm.'' reports per-source normalized-unique malicious identities and therefore need not sum to 7,539 because identities can occur in multiple sources; Main-benign counts precede conflict exclusion.}
\label{tab:core_sources}
\scriptsize
\setlength{\tabcolsep}{3.2pt}
\begin{tabular}{lrrrl}
\toprule
Source & Core raw & Norm. & Main benign & Core provenance \\
\midrule
MalSkillBench & 3,944 & 3,430 & 0 & mixed/unresolved \\
MaliciousAgentSkillsBench & 157 & 157 & 0 & wild \\
SkillLeakBench & 79 & 67 & 0 & wild \\
Agent Skill Malware & 124 & 124 & 0 & wild \\
AgentTrap & 91 & 89 & 50 & test fixture \\
SkillTrojan & 1 & 1 & 2 & backdoored \\
SkillHarm & 879 & 739 & 0 & injected/backdoored \\
SkillTrustBench & 2,863 & 2,861 & 1,643 & injected/synth./wild \\
ATR Skill Security & 32 & 29 & 466 & wild/test fixture \\
SkillFortifyBench & 90 & 90 & 90 & synthetic \\
SkillSafetyBench & 154 & 154 & 0 & injected \\
\bottomrule
\end{tabular}
\end{table}

%% file: sections/05_threat_landscape.tex
\section{Threat Landscape of \benchname}
\label{sec:threat_landscape}

A comprehensive benchmark should characterize both artifact scale and the malicious behavior covered by its contributing sources. We therefore harmonize source-native threat metadata into a common attack taxonomy without inferring labels from Skill text. Direct labels and deterministic strong-semantic mappings support attack characterization for \textbf{4,983 of 7,505 primary malicious identities (66.4\%)} across \textbf{11 multi-label attack categories}. Nine of the 11 Core-contributing sources provide mappable per-unit attack annotations; unsupported units remain unannotated, with no model-inferred labels added.

\begin{figure}[t]
    \centering
    \includegraphics[width=\linewidth]{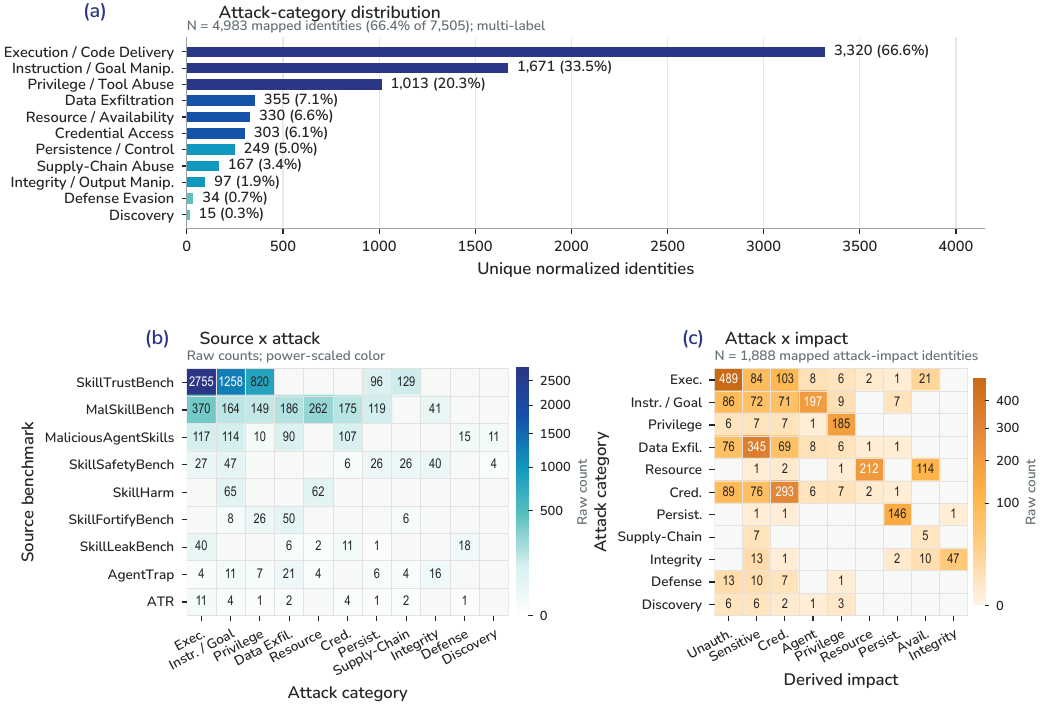}
    \caption{Threat landscape of \benchname. \textbf{(a)} Prevalence of 11 harmonized attack categories among 4,983 malicious identities with direct or strong-semantic source-native mappings; categories are multi-label and percentages use this mapped subset. \textbf{(b)} Source-wise attack composition for the nine Core sources with mappable annotations; cells show raw counts, with power-normalized color intensity for visibility. \textbf{(c)} Attack--impact co-occurrence for the 1,888 identities with both mappings. Impact labels are conservative high-level derived mappings available for a bounded subset of the corpus.}
    \label{fig:threat_landscape}
\end{figure}

\subsection{Attack Coverage Is Broad but Source-Dependent}
\label{sec:threat_attacks}

Figure~\ref{fig:threat_landscape}a shows several recurring behaviors. \emph{Execution / Code Delivery} appears in \textbf{3,320} identities (66.6\% of the mapped subset), followed by \emph{Instruction / Goal / Memory Manipulation} in \textbf{1,671} (33.5\%) and \emph{Privilege / Tool / Authority Abuse} in \textbf{1,013} (20.3\%). Credential access, data exfiltration, persistence, supply-chain abuse, resource abuse, integrity manipulation, defense evasion, and discovery provide additional coverage. Because the taxonomy is multi-label, percentages do not sum to 100\%.

The aggregate distribution hides strong source concentration (Figure~\ref{fig:threat_landscape}b). SkillTrustBench contributes 2,755 of the 3,320 execution-coded identities and 1,258 of the 1,671 instruction-manipulation identities, whereas MalSkillBench contributes 262 of the 330 resource/availability-abuse identities. Other categories draw from different mixtures of wild, synthetic, injected, backdoored, and test-fixture sources. The contributing datasets therefore provide complementary, source-specific threat coverage: different sources can expose substantially different threat profiles.

\subsection{Derived Impacts Provide a More Conservative View}
\label{sec:threat_impacts}

Explicit source-native impact labels are sparse, so impact characterization remains separate from the attack taxonomy. Conservative high-level impact mappings are supported for \textbf{2,128 of 7,505 malicious identities (28.4\%)}, and \textbf{1,888 (25.2\%)} have both attack and impact mappings. Figure~\ref{fig:threat_landscape}c therefore visualizes this bounded mapped subset. Within that subset, execution/code delivery most often co-occurs with unauthorized code execution or system control (489 identities), data exfiltration/disclosure with sensitive-data disclosure (345), and credential access with credential compromise (293). These are descriptive co-occurrences, not causal transitions or benchmark-wide ground truth.

\paragraph{Takeaway.}
The consolidated benchmark broadens observed threat coverage while revealing how unevenly that coverage is distributed across existing resources. This heterogeneity is itself a benchmark finding: no single source represents the full observed malicious-Skill landscape, which motivates both multi-source construction and source-aware evaluation.

%% file: sections/06_detection.tex
\section{Benchmarking Malicious Agent Skill Detection}
\label{sec:evaluation}
\label{sec:results}

\subsection{Evaluation Design}
\label{sec:evaluation_splits}
\label{sec:evaluation_baselines}
All evaluations use the frozen 9,740-unit master table (7,505 malicious / 2,235 benign). \textbf{Random} is label-stratified 70/10/20; \textbf{Malicious-Structural-Disjoint} keeps each of the 4,588 malicious structural families atomic across partitions; and \textbf{Source-Disjoint} holds out SRC009, SRC011, and SRC012 after removing eight identities whose provenance crosses held-out and non-held-out sources. Source-Balanced Random is a composition diagnostic; exact split inventories and leakage audits are in Appendix~\ref{app:evaluation}.

Learned baselines use only inert primary Skill instruction text: word TF--IDF with logistic regression or linear SVM, and character \texttt{char\_wb} TF--IDF with linear SVM. We report Macro-F1, malicious recall, and benign $\mathrm{FPR}_{B}$. The same primary-artifact representation is scanned by three public tools with pre-registered gates: \textbf{Cisco-local-behavioral} (local HIGH/CRITICAL gate), \textbf{SkillFortify-offline} (MEDIUM+), and \textbf{SkillSpector-static} (LLM disabled; native block gate). Technical failures remain abstentions; all three have 100\% coverage on Source-Disjoint. Because learned models are protocol-trained and scanners use fixed external configurations, this is an operational comparison that does not isolate model capacity (Appendix~\ref{app:scanners}).

\subsection{Finding 1: Random Evaluation Overstates Robustness}
\label{sec:results_generalization}
Table~\ref{tab:main_results} reports the learned baselines. Random Macro-F1 is 0.882--0.932. Enforcing malicious structural-family disjointness reduces Random by only 0.016--0.038, whereas Source-Disjoint falls to 0.653--0.665. The strongest word TF--IDF SVM scores \textbf{0.932/0.916/0.665} on Random/Malicious-Structural-Disjoint/Source-Disjoint. Thus controlling broader reuse changes performance modestly relative to the gap created by holding out entire sources.

\input{tables/main_results}

\subsection{Finding 2: The Main Cross-Source Failure Is Benign Over-Flagging}
\label{sec:results_overflagging}
On Source-Disjoint, learned-detector malicious recall remains 94.4--95.7\%, but benign FPR reaches 62.0--64.4\%. The word-SVM flags \textbf{340/545 benign Skills} while missing only \textbf{37/839 malicious Skills}; SRC011 contributes 293 false positives. A format-only model is much weaker (Random Macro-F1 0.486). Balancing lowers word-SVM Source-Disjoint benign FPR from 62.4\% to 43.3\% and raises Macro-F1 from 0.665 to 0.710, but neither balancing nor scaffold sanitization removes the source-conditioned gap (Appendices~\ref{app:results_full} and~\ref{app:robustness}). The learned models therefore fail primarily by treating unfamiliar benign source conventions as malicious, not by losing malicious recall.

\subsection{Finding 3: Existing Scanners Do Not Resolve the Trade-off}
\label{sec:results_scanners}
Table~\ref{tab:scanner_source_disjoint} compares the word-SVM with three fixed scanner configurations on the same Source-Disjoint test. Cisco and SkillSpector keep benign FPR at \textbf{1.1\%} and \textbf{0.55\%}, but detect only \textbf{2.5\%} and \textbf{0\%} of malicious Skills. SkillFortify raises malicious recall to \textbf{25.3\%} while benign FPR reaches \textbf{49.9\%}; the word-SVM reaches 95.6\% recall at 62.4\% FPR.

\input{tables/scanner_source_disjoint}

On SRC011's 455 benign Skills, the word-SVM produces 293 false positives, SkillFortify 272, Cisco 6, and SkillSpector 3. SkillFortify also benefits from the related SRC012/SkillFortifyBench source: excluding SRC012 lowers its Source-Disjoint Macro-F1 from 0.349 to 0.254 and worsens both recall and FPR. This is a source-overlap sensitivity, not evidence of memorization. Across the evaluated methods, lower false-positive rates are therefore obtained only by moving to much lower malicious recall; no detector occupies the desired high-recall, low-FPR regime on held-out sources.

%% file: tables/main_results.tex
\begin{table}[H]
\centering
\caption{Primary detector results. Values are three-seed mean Macro-F1. SB is Source-Balanced Random; $\mathrm{FPR}_{B}$ is the benign false-positive rate (malicious is the positive class).}
\label{tab:main_results}
\small
\setlength{\tabcolsep}{3.6pt}
\begin{tabular}{lrrrrrr}
\toprule
Model & Random & SB & M-Struct. & Source & Rand. $\mathrm{FPR}_{B}$ & Src. $\mathrm{FPR}_{B}$ \\
\midrule
Word TF--IDF + LR  & .882 & .874 & .860 & .661 & .105 & .620 \\
Word TF--IDF + SVM & \textbf{.932} & \textbf{.921} & \textbf{.916} & \textbf{.665} & .094 & .624 \\
Char TF--IDF + SVM & .921 & .907 & .883 & .653 & .098 & .644 \\
\bottomrule
\end{tabular}
\end{table}

%% file: tables/scanner_source_disjoint.tex
\begin{table}[H]
\centering
\caption{Source-Disjoint operational comparison. Scanner coverage is 100\%; scanners use fixed external settings, while word-SVM is protocol-trained.}
\label{tab:scanner_source_disjoint}
\small
\setlength{\tabcolsep}{5.0pt}
\begin{tabular}{llrrr}
\toprule
Detector & Setting & Mal. recall & Benign FPR & Macro-F1 \\
\midrule
Word TF--IDF + SVM & learned & \textbf{.956} & .624 & \textbf{.665} \\
Cisco-local-behavioral & fixed scanner & .025 & .011 & .308 \\
SkillFortify-offline & fixed scanner & .253 & .499 & .349 \\
SkillSpector-static & fixed scanner & .000 & \textbf{.006} & .281 \\
\bottomrule
\end{tabular}
\end{table}

%% file: sections/07_discussion.tex
\section{Discussion}
\label{sec:discussion}

\paragraph{Detection quality is inherently two-sided.}
Source-Disjoint measures source-conditioned stress under coupled provenance, construction, labeling, and class composition. Learned detectors preserve 94.4--95.7\% malicious recall while benign FPR reaches 62.0--64.4\%; Cisco and SkillSpector reduce benign FPR to 1.1\% and 0.55\% but detect only 2.5\% and 0\% of held-out malicious Skills, and SkillFortify lies between these regimes. Random-split Macro-F1, attack recall, or false alarms alone therefore provide incomplete views of detector utility. A realistic benchmark must report malicious detection and benign over-flagging together.

\paragraph{Comprehensiveness requires diversity and control.}
Figure~\ref{fig:threat_landscape} shows that contributing datasets cover different attack profiles, while Section~\ref{sec:results_overflagging} shows that benign diversity is equally consequential. Simply adding more malicious rows is insufficient if those rows duplicate existing content, concentrate in a few construction styles, or are paired with narrow benign data. Identity, structural reuse, attack semantics, and source provenance should remain separate benchmark concepts: identity/conflict control precedes splitting; source/evidence metadata support audit and stratification and are excluded from detector features.

%% file: sections/08_limitations.tex
\section{Limitations and Threats to Validity}
\label{sec:limitations}

\paragraph{Coverage and characterization.}
\emph{Comprehensive} refers to broad, traceable artifact-level consolidation of 13 frozen public sources; coverage is bounded by those sources. Source-Disjoint measures source-conditioned generalization with provenance, construction, labeling, and class composition coupled. SRC002/SRC004 use author-provided historical artifact snapshots, with four unresolved SRC004 cases excluded. Attack mappings cover 4,983/7,505 malicious identities, derived impacts 2,128, and their intersection 1,888; structural families are operational reuse groupings for audit and split control.

\paragraph{Benign and detector scope.}
The Main benign pool favors label confidence over ecological balance, so its source coverage remains narrower than the malicious side. Learned baselines and the common scanner track use static primary Skill artifacts and exclude package-level/runtime behavior. SkillSpector is evaluated in static \texttt{--no-llm} mode; SkillFortify's SRC012 overlap benefits its held-out aggregate; and Cisco uses a documented local compatibility environment with a dependency-version deviation. These conditions bound claims to the evaluated benchmark and configurations.

%% file: sections/09_conclusion.tex
\section{Conclusion}
\label{sec:conclusion}
We introduced \benchname, a comprehensive benchmark that consolidates 13 public sources into 9,740 conflict-clean normalized-unique Skills with explicit identity, reuse, evidence, and threat-coverage accounting. Its 11-category threat characterization shows that existing datasets cover complementary, source-specific attack profiles. Evaluation then reveals a persistent detection trade-off: learned models retain high malicious recall by over-flagging unfamiliar benign Skills, while fixed off-the-shelf scanners reduce false alarms only by moving to much lower recall. Reliable malicious Agent Skill detection therefore remains open, and progress requires diverse malicious and benign coverage together with source-aware, two-sided evaluation.

%% file: sections/10_statements.tex
\section*{Reproducibility Statement}
We freeze source revisions, the canonical metadata manifest, Core eligibility rules, exact and normalized hashes, structural-family assignments, benchmark master tables, split manifests, leakage checks, threat-characterization mappings, text-preprocessing rules, scanner verdict contracts, and detector configurations. The public \href{https://protectskills.github.io/MaliciousSkillBench/}{project page}, \href{https://github.com/protectskills/MaliciousSkillBench}{GitHub repository}, and \href{https://huggingface.co/datasets/ProtectSkills/MaliciousSkillBench}{Hugging Face dataset} document the benchmark release, frozen evaluation protocols, and released artifacts underlying the reported results without requiring execution of untrusted Skill payloads. All 9,740 benchmark identities are publicly represented. Exact frozen static Skill text is available for 9,735 identities (7,500 malicious and 2,235 benign); for five malicious records containing sensitive credential material, the exact original text is withheld and a sanitized representation is provided instead. These five sanitized representations are not bit-for-bit identical to the frozen inputs used in the reported experiments.

\section*{Ethics Statement}
This work studies malicious Agent Skills and therefore involves dual-use security artifacts. We treat all Skill contents as untrusted static data and do not execute malicious payloads, helper scripts, or embedded commands during benchmark construction or the reported detector experiments. Full-package redistribution remains subject to source-specific licenses and constraints. For five malicious records containing sensitive credential material, the exact frozen text is withheld and a sanitized public representation is released instead. The benchmark is intended to support defensive detection and auditing, and released documentation should avoid operationalizing harmful payloads beyond what is necessary for reproducible security research. The detailed release, redistribution, safety, intended-use, and maintenance contract is documented in Appendix~\ref{app:release}.

\section*{AI Use Statement}
Generative AI tools were used to assist literature search, workflow scripting, prose drafting and editing, LaTeX organization, consistency checking, and the explicitly reported blind structural-template review. The blind review is identified in the paper as LLM-based and is not treated as human ground truth. The authors remain responsible for the scientific content and reviewed the reported claims, citations, benchmark statistics, experimental results, and final manuscript text.

%% file: references_manual.tex

%% file: appendix/appendix.tex
\section{Source Registry and Benchmark Scope}
\label{app:sources}

\subsection{Scope and acquisition principles}
\label{app:sources_scope}
\benchname is built from a frozen public-source registry with pinned source revisions. We include public datasets and security corpora that materially characterize the Agent Skill attack surface, even when their released unit or supervision is not eligible for the Core malicious pool. This distinction is deliberate: the registry is broader than the primary detection benchmark. A source can therefore contribute Core malicious artifacts, high-confidence Main-benign artifacts, Auxiliary evidence, or a mixture of these roles.

The public source-of-record revisions are frozen and traceable. We preserve \emph{reported} and \emph{acquired} counts separately and do not synthesize missing records to force agreement with a paper or README. Acquisition is static only: archives and text are inspected for canonicalization and hashing, but Skill code, helper programs, payloads, URLs, installers, evaluators, and test harnesses are never executed. For SRC002 and SRC004, the public revisions remain the source of record while eligible malicious rows are linked to author-provided historical research snapshots that recover the corresponding static Skill artifacts. Table~\ref{tab:app_full_registry} gives the complete 13-source snapshot used by this paper.

\input{tables/appendix_source_registry}

\subsection{Core, Main-benign, and Auxiliary roles}
\label{app:sources_roles}
The Core malicious pool is intentionally narrower than the source registry. A record enters Core only when the released or recovered research unit is a Skill artifact (\texttt{skill\_md} or \texttt{skill\_package}), canonical intent is malicious, static content is available for content hashing/detector input, and the record is traceable to the frozen source revision. \textbf{Eleven sources} satisfy these requirements for at least one malicious artifact. Five sources---SRC006, SRC008, SRC010, SRC011, and SRC012---also contribute strong/moderate Main-benign artifacts before conflict exclusion, yielding 2,251 raw Main-benign records (2,238 exact- and normalized-unique units).

The remaining source content is not discarded. Vulnerability-only records, task-pair attack cases, metadata-only verdicts without recovered artifacts, automated scanner labels, non-Skill formats, weak/silver benign labels, and environment-only cases are retained as Auxiliary data with their original source semantics. Table~\ref{tab:app_auxiliary_only} explains the two registry sources that contribute no Core malicious artifact. Mixed-role sources are handled at record level: SRC002 and SRC004 now contribute recovered malicious Skill artifacts to Core while their public metadata-only/non-malicious rows retain Auxiliary semantics; SkillFortifyBench contributes only Claude Skill artifacts to Core/Main while MCP/OpenClaw formats remain Auxiliary; and SkillSafetyBench contributes 154 Skill-carried attack cases while its single environment-only case remains Auxiliary.

\input{tables/appendix_auxiliary_sources}

\subsection{Frozen revisions and release constraints}
\label{app:sources_revisions}
Reproducibility requires distinguishing the source snapshot used for analysis from source/package redistribution beyond the public benchmark text layer. Table~\ref{tab:app_source_revisions} records the frozen revision, acquisition status, upstream license, and source/package redistribution policy. ``Allowed'' refers only to the acquired public artifact under the recorded license; it does not override upstream terms. For SRC002 and SRC004, the recovered malicious artifacts come from author-provided historical research snapshots; the exact frozen static Skill text for accepted public identities is now released while that historical provenance remains recorded. Where full package/archive redistribution is restricted or unclear, the benchmark exposes the per-identity text representation together with source adapters, revision identifiers, metadata, and hashes, while omitting the underlying package bytes.

\input{tables/appendix_source_revisions}

\subsection{Documented source discrepancies and recovery cases}
\label{app:sources_discrepancies}
We preserve version/count discrepancies and recovery boundaries as explicit provenance. The most consequential cases are:
\begin{itemize}[leftmargin=1.25em,itemsep=2pt,topsep=2pt]
  \item \textbf{SRC002 (MaliciousAgentSkillsBench).} The public snapshot contains 98,380 metadata rows, including 157 confirmed-malicious rows, but does not release the corresponding raw malicious Skill packages. We link all 157 rows to author-provided historical research artifacts with high-confidence mapping and admit those static artifacts to Core. The public metadata remains the source of record for provenance; exact frozen static Skill text for the accepted benchmark identities is included in the public benchmark release, without treating the historical package snapshot as the public source of record.
  \item \textbf{SRC003 (Skill-Inject).} The paper reports 202 injection--task pairs over 23 Skills, whereas the pinned current commit contains 319 pairs over 33 referenced host Skill paths. We retain the pinned 319-row snapshot, mark the acquisition as partial relative to the paper, and keep the task-pair unit outside Core.
  \item \textbf{SRC004 (SkillLeakBench).} The corresponding empirical study reports 520 affected Skills \citep{chen2026credentialleak}; the public snapshot contains 437 vulnerable and 83 malicious metadata rows but does not contain the raw malicious \texttt{SKILL.md} artifacts. Author-provided historical packages resolve 79 malicious artifacts under the Core rule, and exact frozen static Skill text for accepted benchmark identities is included in the public benchmark release. Four ambiguous multi-Skill cases remain outside Core, and the 437 vulnerability-only rows remain Auxiliary.
  \item \textbf{SRC005 (Agent Skill Malware).} The README reports 350 rows (127 malicious + 223 benign), but the fixed revision contains 347 (124 malicious + 223 benign). The three missing malicious records are not reconstructed; Core therefore contains the 124 acquired malicious artifacts.
  \item \textbf{SRC006 (AgentTrap).} A previous metadata-only ingestion did not contain runnable Skill packages. The frozen P1 revision safely recovered all 141 public packages (91 malicious + 50 benign) for local static research. Because the upstream archive provides no explicit redistribution terms, full package/archive redistribution remains restricted; the public benchmark still provides its per-identity static-text representation and reproducibility metadata.
  \item \textbf{SRC008 (SkillTrojan).} The paper reports a curated SkillTrojanX corpus with 3,000+ backdoored Skills, but the pinned official repository exposes only three example Skill packages. We acquire exactly those three examples (one malicious, two benign) and do not generate the missing paper-scale corpus.
  \item \textbf{SRC013 (SkillSafetyBench).} The benchmark reports 155 adversarial cases. Static carrier auditing finds 154 cases with an explicitly injected/modified file inside a Skill package and one environment-only case; the former enter Core and the latter remains Auxiliary.
\end{itemize}

These cases illustrate why \benchname treats source revision, released artifact unit, recovered-artifact provenance, and acquired count as first-class benchmark metadata. ``Paper-reported size'' and ``artifact-level samples available for a frozen evaluation snapshot'' are treated as distinct quantities.

\input{appendix/appendix_taxonomy}

\input{appendix/appendix_reuse}

\input{appendix/appendix_benign}

\input{appendix/appendix_threat}

\input{appendix/appendix_evaluation}

\input{appendix/appendix_experimental}

\input{appendix/appendix_robustness}

\input{appendix/appendix_results}

\input{appendix/appendix_scanners}

\input{appendix/appendix_examples}

\input{appendix/appendix_release}

%% file: tables/appendix_source_registry.tex
\begin{table}[H]
\centering
\caption{Complete frozen source registry used by this paper. ``Reported'' is the source/paper count at the recorded public revision; ``Acquired'' is the number of source records materialized into the canonical registry. Core M is the number of malicious Skill artifacts admitted to Core, including SRC002/SRC004 artifacts recovered from author-provided historical research snapshots. Main B is the raw high-confidence benign contribution before cross-label conflict exclusion. A dash denotes zero.}
\label{tab:app_full_registry}
\scriptsize
\setlength{\tabcolsep}{2.7pt}
\begin{tabularx}{\textwidth}{@{}l>{\raggedright\arraybackslash}p{2.55cm}rrrr>{\raggedright\arraybackslash}X@{}}
\toprule
ID & Source & Reported & Acquired & Core M & Main B & Canonical benchmark role \\
\midrule
SRC001 & MalSkillBench & 7,944 & 7,944 & 3,944 & -- & Core malicious; source benign labels retained outside Main \\
SRC002 & MaliciousAgentSkills\allowbreak Bench & 98,380 & 98,380 & 157 & -- & Recovered malicious artifacts enter Core; public metadata/security labels otherwise remain Auxiliary \\
SRC003 & Skill-Inject & 202 & 319 & -- & -- & Auxiliary only: released unit is an injection--task pair \\
SRC004 & SkillLeakBench & 520 & 520 & 79 & -- & Recovered malicious artifacts enter Core; vulnerability-only and unresolved rows remain Auxiliary \\
SRC005 & Agent Skill Malware & 350 & 347 & 124 & -- & Core malicious; weak benign labels retained outside Main \\
SRC006 & AgentTrap & 141 & 141 & 91 & 50 & Core malicious + Main benign \\
SRC007 & ClawHub Security Signals & 67,453 & 67,453 & -- & -- & Auxiliary only: automated silver scanner labels \\
SRC008 & SkillTrojan & 3,000+ & 3 & 1 & 2 & Core example + Main benign examples; full reported corpus unavailable \\
SRC009 & SkillHarm & 879 & 879 & 879 & -- & Core malicious \\
SRC010 & SkillTrustBench & 5,520 & 5,520 & 2,863 & 1,643 & Core malicious + Main benign; vulnerable rows remain Auxiliary \\
SRC011 & ATR Skill Security Benchmark & 498 & 498 & 32 & 466 & Core malicious + Main benign \\
SRC012 & SkillFortifyBench & 540 & 540 & 90 & 90 & Claude Skill artifacts enter Core/Main; MCP/OpenClaw formats remain Auxiliary \\
SRC013 & SkillSafetyBench & 155 & 155 & 154 & -- & Skill-carried attack cases enter Core; one environment-only case is Auxiliary \\
\bottomrule
\end{tabularx}
\end{table}

%% file: tables/appendix_auxiliary_sources.tex
\begin{table}[H]
\centering
\caption{Why the two registry sources with zero Core malicious artifacts remain Auxiliary. Auxiliary status is a semantic decision about the released unit/evidence, not a judgment that the source is unimportant.}
\label{tab:app_auxiliary_only}
\scriptsize
\setlength{\tabcolsep}{3.2pt}
\begin{tabularx}{\textwidth}{@{}l>{\raggedright\arraybackslash}p{2.45cm}>{\raggedright\arraybackslash}X@{}}
\toprule
ID & Released object / supervision & Reason for Auxiliary-only status \\
\midrule
SRC003 & injection--task pairs over host Skills & The benchmark unit is a task pair with no malicious Skill artifact; contextual cases are additionally retained as harmful/dual-use under their source semantics. \\
SRC007 & sanitized marketplace content with automated security signals & Labels are automated silver supervision (clean/suspicious/malicious signals), so scanner-positive records are not promoted to malicious Core ground truth. \\
\bottomrule
\end{tabularx}
\end{table}

%% file: tables/appendix_source_revisions.tex
\begin{table}[H]
\centering
\caption{Frozen source revisions, licensing, and source/package redistribution policy. Revisions are abbreviated here for readability; the release registry stores full commit/revision identifiers and checksums. The policy column concerns upstream package/archive material beyond the public per-identity static-text layer described in Appendix~\ref{app:release}. ``Partial'' means the pinned public artifact does not exactly reproduce the count or artifact set claimed by the associated paper/README; no missing records are reconstructed.}
\label{tab:app_source_revisions}
\scriptsize
\setlength{\tabcolsep}{2.8pt}
\begin{tabularx}{\textwidth}{@{}ll>{\raggedright\arraybackslash}p{1.85cm}>{\raggedright\arraybackslash}p{2.45cm}>{\raggedright\arraybackslash}X@{}}
\toprule
ID & Revision & Status & Upstream license & Source/package redistribution policy \\
\midrule
SRC001 & \texttt{06e08312} & completed & academic-research-only & adapter only \\
SRC002 & \texttt{422bf340} & completed & MIT & exact frozen static Skill text released; historical snapshot provenance retained \\
SRC003 & \texttt{182f3d9d} & partial & MIT & allowed \\
SRC004 & \texttt{8264436a} & completed & MIT & exact frozen static Skill text released; historical snapshot provenance retained \\
SRC005 & \texttt{5cff435d} & partial & MIT & allowed \\
SRC006 & \texttt{ab13f59d} & completed & terms not provided & metadata/hash only \\
SRC007 & \texttt{69dcbd32} & completed & MIT & allowed \\
SRC008 & \texttt{2864c752} & partial & Apache-2.0 snapshot & allowed for acquired repository snapshot \\
SRC009 & \texttt{cc21f909} & completed & CC-BY-4.0 & metadata/hash only \\
SRC010 & \texttt{f90517b7} & completed & CC-BY-NC-SA-4.0 & metadata/hash only \\
SRC011 & \texttt{7219b10d} & completed & MIT & allowed \\
SRC012 & \texttt{eb9d5a9c} & completed & MIT & allowed \\
SRC013 & \texttt{e0589d7b} & completed & Apache-2.0 & allowed \\
\bottomrule
\end{tabularx}
\end{table}

%% file: appendix/appendix_taxonomy.tex
\section{Canonical Taxonomy and Data Model}
\label{app:taxonomy}

The unified registry is intentionally multi-axis: \emph{what a record is labeled as}, \emph{where it came from}, \emph{how strongly that label is supported}, \emph{what evidence supports it}, and \emph{whether it is an original or derived artifact} are represented separately. This design prevents a source-specific word such as ``malicious,'' ``suspicious,'' or ``clean'' from silently determining benchmark role. The schema is frozen for this paper; source labels are retained losslessly alongside canonical fields.

\subsection{Canonical record schema}
\label{app:taxonomy_schema}
Table~\ref{tab:app_schema_groups} summarizes the field families used by the frozen benchmark. Stable identifiers and source metadata preserve traceability; semantic fields separate intent, provenance, confidence, and evidence; content hashes support conservative identity checks; lineage fields are populated only from source-published relationships; and release fields document what may be redistributed. Fields used only for auditing or split construction are never supplied to the text detectors in Section~\ref{sec:evaluation}.

\input{tables/appendix_schema_groups}

The canonical manifest contains 182,699 source records. Its observed intent distribution is 142,490 \texttt{benign}, 29,840 \texttt{uncertain}, 8,779 \texttt{malicious}, 1,451 \texttt{vulnerable}, and 139 \texttt{harmful\_or\_dual\_use}. Importantly, canonical intent is not benchmark membership: \textbf{8,414} of the 8,779 malicious-intent records satisfy the stricter Core artifact rule in the frozen benchmark, while benign records are further separated into Main and Auxiliary pools according to evidence strength and artifact availability.

\subsection{Intent labels}
\label{app:taxonomy_intent}
Canonical intent answers \emph{what claim is being made about the record}; it does not encode origin or confidence. Table~\ref{tab:app_intent_labels} gives the operational definitions used during source mapping. We choose the most conservative label supported by the source semantics: ambiguous scanner findings remain \texttt{uncertain}, vulnerability findings remain \texttt{vulnerable}, and harmful-but-not-intentionally-malicious cases remain \texttt{harmful\_or\_dual\_use}.

\input{tables/appendix_intent_labels}

\subsection{Provenance taxonomy}
\label{app:taxonomy_provenance}
Provenance answers \emph{how the artifact arose}. It is independent of label confidence: a synthetic artifact can have strong ground truth, while a wild record can have only silver scanner evidence. Table~\ref{tab:app_provenance} reports the controlled vocabulary and the frozen Core-malicious distribution. \texttt{mixed\_unresolved} is used when a source reports an aggregate mixture (for example, wild and generated records) but does not publish a trustworthy row-level mapping. We do not infer such provenance from filenames, content similarity, or repository context.

\input{tables/appendix_provenance}

\subsection{Label strength and evidence are orthogonal}
\label{app:taxonomy_evidence}
We represent confidence and evidentiary mechanism as two separate axes. \emph{Ground-truth strength} describes how strongly the benchmark is willing to rely on a label; \emph{evidence level} describes what kind of evidence produced that label. For example, a deliberately constructed attack can be \texttt{strong}/\texttt{constructed}, a manually curated benign artifact can be \texttt{moderate}/\texttt{static}, and a marketplace record can be \texttt{silver}/\texttt{scanner}. Consequently, evidence type must not be interpreted as an ordinal confidence score.

\input{tables/appendix_label_evidence}

In the frozen registry, the full 182,699-record manifest contains 169,676 silver, 10,525 strong, 2,275 moderate, and 223 weak labels. Scanner evidence dominates the full registry because the largest marketplace-security corpora are silver-standard resources. By contrast, the \textbf{8,414 raw Core malicious artifacts} contain only strong (8,237) or moderate (177) labels, supported by \texttt{constructed} (3,966), \texttt{human+runtime} (4,180), \texttt{static} (177), or \texttt{runtime} (91) evidence. Scanner-only evidence therefore never enters Core malicious ground truth.

\subsection{Artifact units and Core eligibility}
\label{app:taxonomy_artifacts}
Artifact unit records \emph{what the released or recovered row actually represents}. This distinction is central to the benchmark: a source may publish a security-relevant task pair, scanner verdict, MCP configuration, or environment fixture without releasing a malicious Agent Skill artifact that can be hashed and evaluated as a Skill. Table~\ref{tab:app_artifact_units} reports the observed values in the frozen analysis manifest. Only \texttt{skill\_package} and \texttt{skill\_md} are eligible artifact units for the Core malicious pool, and even those must also satisfy intent, traceability, content-availability, and source-mapping requirements.

\input{tables/appendix_artifact_units}

\subsection{Evidence-based lineage semantics}
\label{app:taxonomy_lineage}
Lineage is recorded only when the source publishes a relationship that can be traced without similarity-based inference. We distinguish three concepts. \texttt{lineage\_status} describes whether a record is standalone, derived, or a variant; \texttt{lineage\_resolution} describes whether the source-published linkage is fully resolved, only family-level resolved, or unresolved; and \texttt{base\_skill\_semantics} states what a source's ``base'' identifier actually denotes. Table~\ref{tab:app_lineage} defines these fields.

\input{tables/appendix_lineage}

Among the 8,414 Core records, lineage status is \texttt{variant} for 3,892, \texttt{derived} for 177, \texttt{independent} for 165, and \texttt{unknown} for 4,180. Official lineage is fully resolved for 1,380 Core records and family-level partially resolved for another 35; 6,999 remain unresolved. The 1,415 records with official full/partial linkage expose 448 namespaced base identifiers in the analysis manifest, but these identifiers are deliberately \emph{not} summed as ``448 independent attacks.'' Depending on the source, a base identifier may denote a pre-attack carrier Skill, an upstream Skill identity, a benchmark family, a logical generation seed, or a standalone benchmark artifact. Structural-template families are a separate audit construct and are never converted into lineage or campaign identity.

\subsection{Source-label preservation and conservative mapping}
\label{app:taxonomy_mapping}
The original \texttt{source\_label} is retained alongside canonical intent, strength, and evidence. Mapping is source-specific and follows each release's documented semantics; the same surface word can support different canonical claims across sources. Table~\ref{tab:app_mapping_examples} gives representative mappings that illustrate this policy. A canonical malicious intent still does not imply Core membership unless an inspectable Skill artifact can be mapped to the frozen source record under the Core eligibility rule.

\input{tables/appendix_mapping_examples}

This multi-axis representation is the basis for the benchmark's evidence-preserving consolidation. It preserves disagreements and evidentiary differences as auditable metadata instead of forcing all upstream taxonomies into a single binary label before deduplication, conflict checking, and split construction.

%% file: tables/appendix_schema_groups.tex
\begin{table}[H]
\centering
\caption{Field groups in the frozen canonical schema. Fields shown are representative; source-native labels are retained in addition to the canonical fields.}
\label{tab:app_schema_groups}
\small
\setlength{\tabcolsep}{4pt}
\renewcommand{\arraystretch}{1.12}
\begin{tabularx}{\textwidth}{p{0.17\textwidth} p{0.34\textwidth} X}
\toprule
\textbf{Field group} & \textbf{Representative fields} & \textbf{Purpose} \\
\midrule
Identity & \texttt{canonical\_id}, \texttt{source\_id}, \texttt{source\_record\_id} & Stable benchmark identity while retaining upstream traceability. \\
Source snapshot & \texttt{source\_dataset}, \texttt{repository\_commit}, \texttt{dataset\_snapshot\_date}, paper version & Reconstruct the exact public revision used for acquisition. \\
Artifact / role & \texttt{artifact\_unit}, \texttt{core\_malicious\_skill}, \texttt{auxiliary\_type} & Separate Skill artifacts from task pairs, metadata, fixtures, and other non-Core units. \\
Semantic labels & \texttt{source\_label}, \texttt{intent\_label}, \texttt{provenance}, \texttt{ground\_truth\_strength}, \texttt{evidence\_level} & Preserve source semantics while exposing orthogonal canonical axes. \\
Content identity & \texttt{skill\_sha256}, \texttt{normalized\_skill\_sha256} & Exact and conservative formatting-normalized identity; no semantic equivalence claim. \\
Lineage & \texttt{lineage\_status}, \texttt{lineage\_resolution}, \texttt{lineage\_basis}, \texttt{base\_skill\_id}, \texttt{variant\_id}, \texttt{base\_skill\_semantics} & Preserve only source-supported parent/family relationships and explain what a base identifier means. \\
Release / safety & \texttt{license}, redistribution status, local/static content reference & Record what can be audited or redistributed without executing potentially malicious artifacts. \\
\bottomrule
\end{tabularx}
\end{table}

%% file: tables/appendix_intent_labels.tex
\begin{table}[H]
\centering
\caption{Canonical intent labels. Counts are over all 182,699 source records before Core/Main filtering.}
\label{tab:app_intent_labels}
\small
\setlength{\tabcolsep}{5pt}
\renewcommand{\arraystretch}{1.12}
\begin{tabularx}{\textwidth}{p{0.20\textwidth} r X}
\toprule
\textbf{Intent} & \textbf{Records} & \textbf{Operational meaning} \\
\midrule
\texttt{malicious} & 8,779 & The source supports intentional malicious or unauthorized behavior associated with the record. Core eligibility is checked separately. \\
\texttt{vulnerable} & 1,451 & A weakness, exposure, or exploitable condition is reported, without sufficient evidence that the Skill artifact itself is intentionally malicious. \\
\texttt{harmful\_or\_}\newline\texttt{dual\_use} & 139 & The case can facilitate harmful behavior, but the released case is not treated as an intentionally malicious Skill artifact. \\
\texttt{benign} & 142,490 & The source presents the record as normal/safe/negative; confidence and admissibility as Main benign are represented separately. \\
\texttt{uncertain} & 29,840 & Suspicious, ambiguous, or scanner-derived security signal that is insufficient for a stronger canonical intent claim. \\
\bottomrule
\end{tabularx}
\end{table}

%% file: tables/appendix_provenance.tex
\begin{table}[H]
\centering
\caption{Provenance vocabulary and frozen Core-malicious distribution. Raw counts are source records and sum to 8,414. Normalized-unique counts are identity coverage by provenance and can overlap when the same normalized identity appears in sources with different provenance labels.}
\label{tab:app_provenance}
\small
\setlength{\tabcolsep}{4pt}
\renewcommand{\arraystretch}{1.10}
\begin{tabularx}{\textwidth}{p{0.18\textwidth} r r X}
\toprule
\textbf{Provenance} & \textbf{Core raw} & \textbf{Norm. unique} & \textbf{Meaning} \\
\midrule
\texttt{wild} & 413 & 364 & Collected from a real upstream ecosystem/repository under the source's published semantics; no benchmark-authored attack construction. \\
\texttt{synthetic} & 295 & 295 & Synthetic standalone Skill/test artifact generated or authored for evaluation. \\
\texttt{injected} & 3,460 & 3,352 & Malicious/harmful content is inserted into a pre-existing carrier, host Skill, or task construction. \\
\texttt{backdoored} & 193 & 161 & Benign-looking Skill contains hidden or triggered malicious behavior. \\
\texttt{test\_fixture} & 109 & 107 & Curated or benchmark-authored fixture used as a controlled security test artifact. \\
\texttt{mixed\_}\newline\texttt{unresolved} & 3,944 & 3,430 & Source reports a mixture of origins but lacks a trustworthy row-level provenance mapping. \\
\texttt{unknown} & 0 & 0 & No supported provenance classification is available. \\
\bottomrule
\end{tabularx}
\end{table}

%% file: tables/appendix_label_evidence.tex
\begin{table}[H]
\centering
\caption{Ground-truth strength and evidence-level vocabularies. These axes are intentionally independent. Full-registry counts are shown for observed values.}
\label{tab:app_label_evidence}
\small
\setlength{\tabcolsep}{4pt}
\renewcommand{\arraystretch}{1.08}
\begin{tabularx}{\textwidth}{p{0.15\textwidth} r X}
\toprule
\multicolumn{3}{l}{\textbf{Ground-truth strength}} \\
\midrule
\texttt{strong} & 10,525 & High-confidence source evidence suitable for primary ground truth when artifact/unit constraints are also satisfied. \\
\texttt{moderate} & 2,275 & Meaningful source evidence with weaker certainty than strong; may still enter Core/Main under source-specific rules. \\
\texttt{silver} & 169,676 & Automated/scanner-derived or weakly supervised labels; retained for analysis but not sufficient alone for Core/Main ground truth. \\
\texttt{weak} & 223 & Weak negative/heuristic evidence; Auxiliary only. \\
\texttt{unknown} & 0 & No confidence assignment. \\
\midrule
\multicolumn{3}{l}{\textbf{Evidence level}} \\
\midrule
\texttt{human+}\newline\texttt{runtime} & 4,621 & Human-grounded label together with runtime/behavioral confirmation. \\
\texttt{runtime} & 141 & Behavioral execution/runtime evidence without the combined human+runtime designation. \\
\texttt{static} & 2,359 & Static artifact inspection or curated source evidence without runtime confirmation. \\
\texttt{scanner} & 169,676 & Automated scanner or security-signal output. \\
\texttt{constructed} & 5,902 & Label is known from controlled benchmark construction, injection, or backdoor generation. \\
\texttt{human}, \texttt{llm}, \texttt{unknown} & 0 & Controlled vocabulary values reserved for explicit human-only, LLM-only, or unavailable evidence; not observed in the frozen analysis manifest. \\
\bottomrule
\end{tabularx}
\end{table}

%% file: tables/appendix_artifact_units.tex
\begin{table}[H]
\centering
\caption{Observed artifact units in the frozen analysis manifest. Core counts show the stricter malicious-Skill subset; recovered SRC002/SRC004 malicious artifacts account for the increase in \texttt{skill\_md} units relative to the earlier snapshot.}
\label{tab:app_artifact_units}
\small
\setlength{\tabcolsep}{4pt}
\renewcommand{\arraystretch}{1.10}
\begin{tabularx}{\textwidth}{p{0.22\textwidth} r r X}
\toprule
\textbf{Artifact unit} & \textbf{All records} & \textbf{Core malicious} & \textbf{Interpretation / Core policy} \\
\midrule
\texttt{skill\_package} & 82,094 & 7,932 & Complete or package-level Skill artifact; Core-eligible when all other rules pass. \\
\texttt{skill\_md} & 1,261 & 482 & Primary Skill instruction artifact; Core-eligible when all other rules pass. \\
\texttt{task\_pair} & 319 & 0 & Attack/task evaluation pair; no released malicious Skill artifact; Auxiliary. \\
\texttt{mcp\_config} & 180 & 0 & MCP-format configuration from a multi-format testbed; outside current Skill Core. \\
\texttt{openclaw\_}\newline\texttt{manifest} & 180 & 0 & OpenClaw-format manifest from a multi-format testbed; outside current Skill Core. \\
\texttt{environment\_}\newline\texttt{fixture} & 1 & 0 & Attack is carried only in the environment and is absent from the Skill package; Auxiliary. \\
\texttt{other} & 98,664 & 0 & Metadata/scanner/vulnerability or other non-Core record representation. \\
\bottomrule
\end{tabularx}
\end{table}

%% file: tables/appendix_lineage.tex
\begin{table}[H]
\centering
\caption{Lineage semantics in the frozen schema. Lineage is source-evidenced and is never inferred from textual similarity or structural clustering.}
\label{tab:app_lineage}
\small
\setlength{\tabcolsep}{4pt}
\renewcommand{\arraystretch}{1.08}
\begin{tabularx}{\textwidth}{p{0.20\textwidth} p{0.24\textwidth} X}
\toprule
\textbf{Field} & \textbf{Values} & \textbf{Meaning} \\
\midrule
\texttt{lineage\_status} & \texttt{independent}, \texttt{derived}, \texttt{variant}, \texttt{unknown} & Logical relationship asserted by the source: standalone identity, explicit derivative, member of a published variant family, or unresolved. \\
\texttt{lineage\_}\newline\texttt{resolution} & \texttt{resolved\_official}; \newline\texttt{partially\_}\newline\texttt{resolved\_official}; \texttt{unresolved} & Whether an official parent/upstream/logical identity is fully available, only a family-level identity is available, or no trustworthy row-level link exists. \\
\texttt{lineage\_basis} & official parent/family ID, upstream slug, benchmark-authored identity, logical construction ID, none & What source-published signal justifies the linkage; filenames and similarity are not accepted as lineage evidence. \\
\texttt{base\_skill\_}\newline\texttt{semantics} & pre-attack carrier, upstream identity, benchmark family, generation seed, standalone identity, unknown & Explains what the source-specific \texttt{base\_skill\_id} denotes so heterogeneous base IDs are not mistaken for a common attack-family unit. \\
\texttt{base\_skill\_id}, \texttt{variant\_id} & source-scoped strings or null & Retain the publisher's identifier when available; values are namespaced/source-specific and are not globally additive. \\
\bottomrule
\end{tabularx}
\end{table}

%% file: tables/appendix_mapping_examples.tex
\begin{table}[H]
\centering
\caption{Representative source-label mappings. Mapping depends on source semantics, evidence, and artifact availability, not on the literal label string alone.}
\label{tab:app_mapping_examples}
\scriptsize
\setlength{\tabcolsep}{3.3pt}
\renewcommand{\arraystretch}{1.10}
\begin{tabularx}{\textwidth}{p{0.10\textwidth} p{0.15\textwidth} p{0.18\textwidth} p{0.18\textwidth} X}
\toprule
\textbf{Source} & \textbf{Source label} & \textbf{Canonical intent} & \textbf{Strength / evidence} & \textbf{Benchmark interpretation} \\
\midrule
SRC002 & \texttt{safe} & \texttt{benign} & silver / scanner & Metadata-level negative signal; retained in Auxiliary and excluded from Main benign. \\
SRC002 & confirmed malicious & \texttt{malicious} & strong / human+runtime & Eligible for Core only after high-confidence linkage to an author-provided historical Skill artifact. \\
SRC007 & \texttt{malicious} & \texttt{uncertain} & silver / scanner & Even a literal ``malicious'' scanner verdict remains uncertain without stronger ground truth. \\
SRC003 & \texttt{contextual} & \texttt{harmful\_or\_}\newline\texttt{dual\_use} & moderate / constructed & Context-dependent harmful task case; not a Core malicious Skill artifact. \\
SRC010 & \texttt{suspicious} & \texttt{vulnerable} & strong/moderate; constructed/static & Source semantics describe vulnerable/suspicious cases without asserting intentional malicious artifacts. \\
SRC004 & \texttt{malicious} & \texttt{malicious} & strong / human+runtime & 79 source-malicious rows map to recovered historical Skill artifacts and enter Core; four ambiguous multi-Skill rows remain outside Core. \\
SRC001 & \texttt{malware} & \texttt{malicious} & strong / human+runtime & Inspectable Skill artifact with supported malicious intent; eligible for Core. \\
SRC010 & \texttt{normal} & \texttt{benign} & moderate/static or strong/constructed & High-confidence benign Skill artifact; eligible for the Main-benign candidate pool. \\
\bottomrule
\end{tabularx}
\end{table}

%% file: appendix/appendix_reuse.tex
\section{Deduplication and Structural Reuse}
\label{app:reuse}

The benchmark separates three increasingly permissive notions of reuse: byte-level identity, conservative normalized-text identity, and static structural-template similarity. These levels serve different purposes and are never collapsed into a single notion of ``same attack.'' Exact and normalized hashes support content identity checks; structural families are an operational grouping used for audit and split construction. No Skill, helper, payload, URL, or executable content is run during any stage of this analysis.

\subsection{Exact and normalized content identity}
\label{app:reuse_identity}
Exact identity is SHA-256 over the acquired Skill content. Normalized identity hashes a deterministic text normalization that is intentionally conservative. The frozen normalization decodes valid UTF-8, removes one leading BOM, converts CRLF/CR to LF, strips trailing spaces/tabs on each line, trims leading/trailing blank lines, collapses runs of three or more blank lines to two, and writes exactly one final LF for non-empty text. It does \emph{not} lowercase text, remove Markdown front matter, code, punctuation, or semantic content, and it does not paraphrase or rewrite instructions. Normalized equality therefore captures formatting-equivalent identity; semantic equivalence is outside the scope of this identity rule.

Applying the two identity levels to \textbf{8,414 Core malicious records} yields \textbf{7,562 exact-unique} and \textbf{7,539 normalized-unique} contents. Exact deduplication removes 852 redundant raw records (10.1\% of Core), while normalization merges 23 additional global identities. Table~\ref{tab:app_dedup_by_source} shows where within-source duplication occurs. Per-source unique counts do not sum to the global unique totals because the same content can appear in more than one source.

\input{tables/appendix_dedup_by_source}

At the exact-hash level, 131 duplicate clusters contain 983 raw Core records, with a largest cluster of 54 records; 28 exact clusters contain at least ten records. Duplication remains concentrated in a few sources, while the recovered SRC004 artifacts add 12 within-source exact duplicates. These counts describe repeated released/recovered artifacts, not independent attack lineage.

\subsection{Cross-source content reuse}
\label{app:reuse_cross_source}
Content reuse also crosses nominal dataset boundaries. Across the frozen Core pool, \textbf{105 exact hashes} and \textbf{119 normalized hashes} occur in more than one source. Those shared hashes involve 830 raw Core records at exact identity and 871 raw Core records at normalized identity. Table~\ref{tab:app_cross_source_content} lists every source pair with non-zero normalized overlap.

\input{tables/appendix_cross_source_content}

The overlap is strongly asymmetric. The 90 normalized hashes shared by SRC001 and SRC005 cover 15.14\% of SRC001 raw Core records but 72.58\% of SRC005. The recovered sources add further observable overlap: 21 normalized hashes are shared by SRC001 and SRC004, covering 35.44\% of SRC004 raw Core records, while three are shared by SRC001 and SRC002. A single normalized identity can occur in more than two sources, so pairwise overlap counts are not additive. We therefore report overlap at the artifact-reuse level and make no inference about source ancestry, common actor identity, or campaign membership.

\subsection{Static structural-similarity pipeline}
\label{app:reuse_structural_method}
Exact/normalized equality does not capture templated variants that preserve a common document scaffold while changing names, endpoints, paths, identifiers, parameters, or local insertions. We therefore maintain a separate static structural-similarity partition over one representative per normalized hash. Cluster membership is determined entirely by fixed text features and the pre-specified medium threshold 0.68; no LLM or embedding model decides membership.

Template normalization conservatively replaces front-matter Skill names, URLs, domains, email/IP values, UUIDs and long IDs, filesystem paths, obvious secret values, long numeric payload parameters, environment variables, and assignment-left-hand-side identifiers. Behavior-bearing terms---including credential access, file reads, network sends, shell/subprocess operations, \texttt{eval}, encryption, download, and execution---are retained. Template similarity is the weighted score
\[
0.36J_{3\text{-gram}} + 0.28C_{\mathrm{tfidf}} + 0.14J_{\mathrm{struct}} + 0.12J_{\mathrm{behavior}} + 0.10\,\mathrm{containment},
\]
where the terms denote token-3-gram Jaccard, TF-IDF cosine, structural-line Jaccard, behavior-token Jaccard, and token containment, respectively.

The frozen partition applies the same operational threshold to \textbf{7,539 normalized Core identities}. Relative to the inherited 7,340-node partition, 199 newly recovered normalized identities are integrated with explicit hash-origin bookkeeping: 25 join inherited families, two form one new recovered-only two-node family, and 172 become new singletons. A full family crosswalk confirms that none of the 4,415 inherited families is split or merged. Table~\ref{tab:app_structural_partition_delta} reports the resulting partition evolution. The 0.68 threshold is used solely as an operational structural-reuse control. It was fixed independently of detector performance and carries no semantic attack interpretation.

\input{tables/appendix_threshold_sensitivity}

\subsection{Family distribution and cross-source structural reuse}
\label{app:reuse_structural_distribution}
At threshold 0.68, the frozen partition contains \textbf{4,588 operational structural-family identifiers}: 3,219 singletons and 1,369 non-singletons containing 4,320 normalized identities. These identifiers are frozen before final cross-label exclusion: \textbf{4,575} are represented by at least one final primary malicious identity, while 13 retained identifiers have no final primary member after conservative cross-label exclusion. Among non-singleton families, median size is 2 and mean size is 3.16; the largest family contains 146 normalized identities spanning SRC001, SRC009, and SRC013. \textbf{113 families} contain artifacts associated with more than one source. Table~\ref{tab:app_cross_source_structural} reports the largest pairwise structural overlaps.

\input{tables/appendix_cross_source_structural}

Structural overlap can be substantial even when exact/normalized hashes do not overlap. SRC009 and SRC013, for example, share no exact or normalized Core hash, yet 26 operational structural families span both sources; those families cover 42.35\% of SRC009 normalized units and 62.99\% of SRC013. The recovered SRC004 artifacts also expose substantial structural reuse with existing corpora: 30 families span SRC001/SRC004, covering 49.25\% of SRC004 normalized identities. This provides scaffold-level reuse evidence under the fixed static method; it does not support claims about copying, shared actors, shared campaigns, or independent attack identity.

\subsection{Connected-component bridge audit}
\label{app:reuse_bridge}
Connected components can link members indirectly through intermediate nodes even when a member is below threshold relative to a selected representative. The frozen summary flags \textbf{15 of 4,588 families} (0.33\%) for possible chaining; the largest 146-unit family is among them. We retain the connected-component family IDs as the operational partition and report the bridge flag explicitly; each family remains a connected component and need not form a clique. The recovery crosswalk additionally confirms that artifact recovery did not merge or split any inherited family.

\subsection{Blind structural positive validation}
\label{app:reuse_blind_validation}
Before the artifact-recovery extension, we performed a separate blind positive check on the inherited structural partition using an independent LLM reviewer on anonymized static artifact pairs. The reviewer received neither cluster IDs, similarity scores, threshold values, sampling categories, algorithm predictions, nor earlier reviewer labels. All 72 cases were valid blind cases; there were no blindness violations or unreadable pairs. The reviewer labeled all 72 pairs as sharing the same structural template, all at high confidence. No embedded instruction, script, helper, URL, or payload was executed or followed.

\input{tables/appendix_blind_structural_validation}

Because the final partition preserves every inherited family without split or merge, this validation remains a bounded positive-coherence check for the sampled inherited within-family pairs. Its coverage is limited to those sampled inherited pairs and leaves the 199 recovered identities, newly created families, clustering recall, and threshold semantics unvalidated. The \textbf{4,588 families} serve as an operational structural partition for reducing scaffold reuse across evaluation splits; they carry no gold attack-ontology interpretation.

\subsection{Interpretation boundaries}
\label{app:reuse_boundaries}
The three reuse levels answer different questions. Exact hashes identify identical acquired content; normalized hashes identify conservatively formatting-equivalent content; structural families identify static scaffold similarity under the frozen feature/threshold pipeline. None of these establishes lineage, actor identity, uploader identity, campaign membership, or attack-mechanism equivalence. Candidate-generation heuristics may miss deep paraphrases that share no selected feature, while static similarity may over-link boilerplate. These limitations motivate the paper's conservative use of structural family IDs as grouping variables for audit and Malicious-Structural-Disjoint evaluation, with no ground-truth attack-label semantics.

%% file: tables/appendix_dedup_by_source.tex
\begin{table}[H]
\centering
\caption{Frozen Core malicious deduplication by source. ``Raw$-$Exact'' is the within-source reduction from repeated byte-identical artifacts; ``Exact$-$Norm.'' is the additional within-source reduction after conservative text normalization. Per-source unique counts do not sum to the global unique totals because cross-source identities are shared.}
\label{tab:app_dedup_by_source}
\scriptsize
\setlength{\tabcolsep}{4.4pt}
\begin{tabular}{@{}lrrrrr@{}}
\toprule
Source & Raw & Exact unique & Norm. unique & Raw$-$Exact & Exact$-$Norm. \\
\midrule
SRC001 & 3,944 & 3,430 & 3,430 & 514 & 0 \\
SRC002 & 157 & 157 & 157 & 0 & 0 \\
SRC004 & 79 & 67 & 67 & 12 & 0 \\
SRC005 & 124 & 124 & 124 & 0 & 0 \\
SRC006 & 91 & 89 & 89 & 2 & 0 \\
SRC008 & 1 & 1 & 1 & 0 & 0 \\
SRC009 & 879 & 739 & 739 & 140 & 0 \\
SRC010 & 2,863 & 2,861 & 2,861 & 2 & 0 \\
SRC011 & 32 & 31 & 29 & 1 & 2 \\
SRC012 & 90 & 90 & 90 & 0 & 0 \\
SRC013 & 154 & 154 & 154 & 0 & 0 \\
\midrule
Global unique & 8,414 & 7,562 & 7,539 & 852 & 23 \\
\bottomrule
\end{tabular}
\end{table}

%% file: tables/appendix_cross_source_content.tex
\begin{table}[H]
\centering
\caption{Non-zero cross-source Core content overlap in the frozen benchmark. Coverage is the fraction of each source's raw Core records participating in a shared normalized hash with the paired source. Pairwise counts overlap when one identity occurs in three or more sources.}
\label{tab:app_cross_source_content}
\scriptsize
\setlength{\tabcolsep}{3.2pt}
\begin{tabular}{@{}llrrrr@{}}
\toprule
Source A & Source B & Exact hashes & Norm. hashes & Norm. cov. A & Norm. cov. B \\
\midrule
SRC001 & SRC002 & 3 & 3 & 0.13\% & 1.91\% \\
SRC001 & SRC004 & 21 & 21 & 0.79\% & 35.44\% \\
SRC001 & SRC005 & 69 & 90 & 15.14\% & 72.58\% \\
SRC001 & SRC006 & 1 & 1 & 0.03\% & 1.10\% \\
SRC001 & SRC010 & 78 & 78 & 14.63\% & 2.72\% \\
SRC001 & SRC011 & 6 & 8 & 0.35\% & 28.13\% \\
SRC004 & SRC005 & 0 & 3 & 7.59\% & 2.42\% \\
SRC004 & SRC009 & 1 & 1 & 1.27\% & 3.53\% \\
SRC004 & SRC010 & 7 & 7 & 13.92\% & 0.24\% \\
SRC004 & SRC011 & 4 & 5 & 6.33\% & 15.63\% \\
SRC005 & SRC010 & 63 & 68 & 54.84\% & 2.38\% \\
SRC005 & SRC011 & 4 & 4 & 3.23\% & 15.63\% \\
SRC010 & SRC011 & 0 & 1 & 0.03\% & 3.13\% \\
\bottomrule
\end{tabular}
\end{table}

%% file: tables/appendix_threshold_sensitivity.tex
\begin{table}[H]
\centering
\caption{Structural-partition evolution at the frozen medium threshold 0.68. The final partition preserves every inherited pre-recovery family without split/merge and integrates 199 newly recovered normalized identities under the same operational similarity rule.}
\label{tab:app_structural_partition_delta}
\scriptsize
\setlength{\tabcolsep}{4.3pt}
\begin{tabular}{@{}lrrrrrr@{}}
\toprule
Snapshot & Nodes & Families & Singletons & Non-singleton & Largest & Cross-source \\
\midrule
Pre-recovery inherited base & 7,340 & 4,415 & 3,059 & 1,356 & 146 & 91 \\
Final frozen partition & 7,539 & 4,588 & 3,219 & 1,369 & 146 & 113 \\
\bottomrule
\end{tabular}
\end{table}

%% file: tables/appendix_cross_source_structural.tex
\begin{table}[H]
\centering
\caption{Largest pairwise cross-source structural overlaps in the frozen partition at threshold 0.68. Coverage is computed over normalized-unique Core units associated with each source. These entries report operational template overlap only, with no lineage or campaign attribution.}
\label{tab:app_cross_source_structural}
\scriptsize
\setlength{\tabcolsep}{3.4pt}
\begin{tabular}{@{}llrrr@{}}
\toprule
Source A & Source B & Shared families & Coverage A & Coverage B \\
\midrule
SRC001 & SRC010 & 39 & 3.38\% & 3.04\% \\
SRC001 & SRC005 & 36 & 3.09\% & 75.81\% \\
SRC001 & SRC004 & 30 & 2.22\% & 49.25\% \\
SRC005 & SRC010 & 29 & 68.55\% & 2.48\% \\
SRC009 & SRC013 & 26 & 42.35\% & 62.99\% \\
SRC004 & SRC010 & 17 & 26.87\% & 1.47\% \\
SRC004 & SRC005 & 13 & 19.40\% & 40.32\% \\
SRC001 & SRC006 & 11 & 1.14\% & 22.47\% \\
SRC001 & SRC002 & 10 & 0.61\% & 8.28\% \\
SRC001 & SRC011 & 9 & 0.35\% & 34.48\% \\
\bottomrule
\end{tabular}
\end{table}

%% file: tables/appendix_blind_structural_validation.tex
\begin{table}[H]
\centering
\caption{Blind structural positive-validation summary. The reviewer saw anonymized static artifact pairs only; cluster identifiers, similarity values, thresholds, sample categories, algorithm predictions, and prior labels were hidden.}
\label{tab:app_blind_structural_validation}
\scriptsize
\setlength{\tabcolsep}{8pt}
\begin{tabular}{@{}lr@{}}
\toprule
Measure & Count \\
\midrule
Total / valid blind cases & 72 / 72 \\
Blindness violations & 0 \\
Same structural template & 72 \\
Different structural template & 0 \\
Uncertain & 0 \\
High-confidence judgments & 72 \\
Unreadable / missing & 0 \\
\bottomrule
\end{tabular}
\end{table}

%% file: appendix/appendix_benign.tex
\section{Benign Pool and Cross-Label Conflicts}
\label{app:benign}

The negative class is constructed conservatively because ``benign,'' ``clean,'' ``not flagged,'' and ``safe'' do not carry the same evidentiary meaning across the 13 sources. The primary detection benchmark therefore uses a high-confidence \emph{Main-benign} pool and keeps weaker or non-comparable negative signals in an \emph{Auxiliary-benign} layer. This appendix documents the candidate gates, deduplication accounting, and the cross-label conflict policy that converts 48,217 benign-side candidates into the 2,235 normalized-unique benign units used by the frozen primary benchmark.

\subsection{Main versus Auxiliary benign candidates}
\label{app:benign_pools}
A benign-side record is eligible for the Main pool only when it is an inspectable Skill artifact, its source semantics support canonical benign intent, its label strength is strong or moderate, static content is available for hashing and detector input, and the record remains traceable to a frozen source revision. Scanner-clean, marketplace-unflagged, silver, weak, metadata-only, and otherwise non-comparable negative signals remain Auxiliary. Table~\ref{tab:app_benign_eligibility} summarizes this gate.

\input{tables/appendix_benign_eligibility}

Applying the gate yields 2,251 raw Main-benign artifacts and 45,966 raw Auxiliary-benign records. The two pools are kept separate throughout benchmark construction; Auxiliary negatives are not silently promoted merely because they are numerous. Table~\ref{tab:app_benign_pool_counts} reports the frozen identity counts.

\input{tables/appendix_benign_pool_counts}

The Main pool is intentionally much smaller than the full benign-side registry. This trades coverage for label comparability: the primary benchmark asks whether a detector can separate Core malicious Skills from higher-confidence benign Skill artifacts, while the much larger Auxiliary layer remains available for future stress tests, scanner-comparison studies, or separately versioned evaluations.

\subsection{Main-benign source contributions}
\label{app:benign_sources}
Only five frozen sources contribute Main-benign candidates: SRC006, SRC008, SRC010, SRC011, and SRC012. Their raw contributions sum exactly to 2,251 (Table~\ref{tab:app_main_benign_sources}). Sources with weaker negative semantics remain outside Main even when they contain records described upstream as benign or clean. For example, SRC001 source-benign records are retained outside Main under the frozen adapter, SRC005 benign rows are weak evidence under the benchmark taxonomy, and scanner-derived negative signals from SRC002/SRC007 remain silver Auxiliary supervision.

\input{tables/appendix_main_benign_sources}

The resulting source mix is a high-confidence reference set assembled from public sources that expose both an inspectable Skill artifact and sufficiently strong benign-side evidence under the frozen mapping rules. Its coverage is bounded by those source and evidence gates. This restriction is also why source and class composition remain entangled in the Source-Disjoint evaluation.

\subsection{Benign deduplication and static handling}
\label{app:benign_dedup}
The benign pools use the same exact and conservative normalized identity definitions as Appendix~\ref{app:reuse}. Main-benign deduplication reduces 2,251 raw records to 2,238 exact-unique contents; normalized text identity introduces no additional merge, leaving 2,238 normalized-unique Main candidates before conflict exclusion. The Auxiliary pool reduces from 45,966 raw records to 44,329 exact-unique and 44,327 normalized-unique contents. Thus the primary benign reference set is already close to content-unique before cross-label auditing, whereas the Auxiliary layer contains substantially more repeated content.

Five benign artifacts contain malformed UTF-8. They are processed statically using replacement decoding for text analysis while preserving raw-byte hashes for identity accounting. No affected artifact is executed, and the decoding choice does not alter the frozen raw hash. Benign structural clustering was not part of the frozen malicious-family pipeline; consequently, Malicious-Structural-Disjoint treats each normalized-unique benign unit as a singleton group and performs no post-freeze benign family inference.

\subsection{Cross-label conflict definition and accounting}
\label{app:benign_conflicts}
After malicious and benign-side canonicalization, we compare exact and normalized content identities across opposite labels. A \emph{cross-label conflict} occurs when the same content identity is represented on the malicious side and on the benign side of the frozen registry. This is a content-level inconsistency signal; it is not itself evidence that either source is wrong.

The frozen audit finds \textbf{34 exact-hash conflicts} and \textbf{34 normalized-hash conflicts}, affecting 162 Core-malicious raw records, four Main-benign raw records, and 39 Auxiliary-benign raw records. The exact and normalized audits expose the same number of conflicting identity groups in this snapshot; conservative normalization does not increase the conflict count. Table~\ref{tab:app_conflict_accounting} shows how those groups affect the primary benchmark.

\input{tables/appendix_conflict_accounting}

The benign-side conflict audit is broader than Main-negative membership. Although 43 benign-side raw records participate across Main and Auxiliary pools, only three normalized Main-benign candidates would otherwise enter the primary benchmark. All \textbf{34 conflicting Core-malicious normalized identities} are excluded under the conservative conflict rule. The pre-conflict 7,539 malicious plus 2,238 Main-benign normalized units therefore lose 37 benchmark units in total, yielding the frozen \textbf{9,740-unit primary table: 7,505 malicious and 2,235 benign}.

\subsection{Conservative conflict exclusion}
\label{app:benign_conflict_policy}
Every conflicting normalized-content group is removed from the primary detection benchmark and retained as a separate label-consistency audit layer. Automatic relabeling by source majority, evidence ranking, filename, marketplace reputation, structural family, or detector score would introduce a new ground-truth judgment that is absent from the upstream releases. This policy is deliberately symmetric at the benchmark level: conflicting content cannot be used as a clean positive or negative training/test unit simply because one upstream source is preferred.

The policy has two consequences. First, the primary benchmark is easier to interpret because no normalized content is knowingly assigned both detection labels. Second, the excluded groups remain scientifically useful: they identify concrete places where public corpora disagree and can support a later human adjudication study. Any such adjudication should be separately versioned so the frozen benchmark used in this paper remains unchanged.

For safety and licensing reasons, the paper does not reproduce full malicious payload text from conflict groups. The audit is defined at the canonical-ID, source-ID, evidence-metadata, and content-hash level; potentially executable content remains inert and subject to the source-specific redistribution constraints in Appendix~\ref{app:sources_revisions}.

\subsection{Interpretation boundaries}
\label{app:benign_boundaries}
The 34 conflicts quantify exact/normalized identity disagreement across corpora with different collection procedures, supervision regimes, and label semantics; the audit does not estimate annotation error rates. Main benign consists of records that satisfy the frozen primary-negative gate, while the remaining 45,966 Auxiliary records retain weaker or non-comparable evidence. The final 2,235-unit benign class is a conservative reference set for the paper's primary detection experiments, with coverage bounded by these source and evidence criteria.

%% file: tables/appendix_benign_eligibility.tex
\begin{table}[H]
\centering
\caption{Frozen eligibility gate for the benign reference pools. ``Auxiliary'' preserves security-relevant negative signals without promoting them to primary benign ground truth.}
\label{tab:app_benign_eligibility}
\scriptsize
\setlength{\tabcolsep}{3.2pt}
\begin{tabularx}{\textwidth}{@{}p{2.5cm}p{5.0cm}X@{}}
\toprule
Criterion & Main-benign requirement & Auxiliary treatment \\
\midrule
Released unit & Inspectable Agent Skill artifact usable for static hashing/detector input & Metadata-only, scanner-only, non-Skill, or otherwise non-comparable negative records remain Auxiliary \\
Canonical intent & \texttt{benign} under source-specific mapping & Uncertain, vulnerable, harmful/dual-use, or unresolved semantics are never collapsed into benign \\
Label strength & Strong or moderate & Silver, weak, ``not flagged,'' marketplace-unflagged, and analogous weak negatives remain Auxiliary \\
Content availability & Static primary Skill content available & Records lacking inspectable Skill text are excluded from the primary negative class \\
Traceability & Frozen source revision and source record can be reconstructed & Untraceable or revision-ambiguous records are not promoted to Main \\
\bottomrule
\end{tabularx}
\end{table}

%% file: tables/appendix_benign_pool_counts.tex
\begin{table}[H]
\centering
\caption{Frozen benign-side pool counts before cross-label conflict exclusion. Main is the primary high-confidence negative pool; Auxiliary retains weaker or non-comparable benign-side evidence.}
\label{tab:app_benign_pool_counts}
\scriptsize
\setlength{\tabcolsep}{5.0pt}
\begin{tabular}{@{}lrrrl@{}}
\toprule
Pool & Raw & Exact-unique & Normalized-unique & Primary benchmark use \\
\midrule
Main benign & 2,251 & 2,238 & 2,238 & Eligible before conflict exclusion \\
Auxiliary benign & 45,966 & 44,329 & 44,327 & Not used as primary negative ground truth \\
\bottomrule
\end{tabular}
\end{table}

%% file: tables/appendix_main_benign_sources.tex
\begin{table}[H]
\centering
\caption{Sources contributing raw Main-benign candidates before deduplication and conflict exclusion. Counts sum to 2,251.}
\label{tab:app_main_benign_sources}
\scriptsize
\setlength{\tabcolsep}{3.2pt}
\begin{tabularx}{\textwidth}{@{}lp{3.1cm}rX@{}}
\toprule
ID & Source & Raw Main B & Frozen benchmark interpretation \\
\midrule
SRC006 & AgentTrap & 50 & Inspectable benign Skill packages with strong/moderate source evidence \\
SRC008 & SkillTrojan & 2 & Benign example packages in the pinned public repository \\
SRC010 & SkillTrustBench & 1,643 & High-confidence source-normal Skill artifacts eligible under the canonical benign mapping \\
SRC011 & ATR Skill Security Benchmark & 466 & Inspectable benign Skill artifacts admitted by the frozen source adapter \\
SRC012 & SkillFortifyBench & 90 & Benign Claude Skill artifacts; non-Skill MCP/OpenClaw formats remain Auxiliary \\
\midrule
Total & & 2,251 & \\
\bottomrule
\end{tabularx}
\end{table}

%% file: tables/appendix_conflict_accounting.tex
\begin{table}[H]
\centering
\caption{Cross-label conflict accounting in the frozen benchmark. Raw affected counts measure source records participating in a conflict; removed-unit counts measure normalized-unique units that would otherwise enter the primary benchmark.}
\label{tab:app_conflict_accounting}
\scriptsize
\setlength{\tabcolsep}{4.4pt}
\begin{tabularx}{\textwidth}{@{}Xrr@{}}
\toprule
Quantity & Count & Benchmark effect \\
\midrule
Exact conflict hashes & 34 & Identity groups with both malicious- and benign-side labels \\
Normalized conflict hashes & 34 & Frozen conflict groups used for exclusion \\
Core-malicious raw records affected & 162 & Source records participating on the malicious side \\
Main-benign raw records affected & 4 & High-confidence negative records participating in conflicts \\
Auxiliary-benign raw records affected & 39 & Auxiliary negative records participating in conflicts \\
Core-malicious normalized units removed & 34 & 7,539 $\rightarrow$ 7,505 \\
Main-benign normalized units removed & 3 & 2,238 $\rightarrow$ 2,235 \\
Total primary units removed & 37 & 9,777 pre-conflict candidates $\rightarrow$ 9,740 final units \\
\bottomrule
\end{tabularx}
\end{table}

%% file: appendix/appendix_threat.tex
\section{Threat Characterization Details}
\label{app:threat}

Section~\ref{sec:threat_landscape} uses source-supported threat metadata to describe what \benchname covers. This appendix records the characterization boundary, final harmonized categories, and supplementary views used to audit that summary. The unit throughout is the normalized-unique primary malicious identity unless stated otherwise. The denominator is therefore the 7,505 final primary malicious identities; raw Core and pre-conflict counts are excluded from these prevalence calculations.

\subsection{Characterization boundary and source-native coverage}
\label{app:threat_boundary}

Threat metadata are heterogeneous across the contributing resources, so we preserve source-native dimensions before harmonization and map only supported semantics into the common taxonomy. Table~\ref{tab:app_threat_metadata_coverage} reports the resulting coverage. Source-native behavior is available for 5,456 identities (72.7\%), attack type/tactic for 4,147 (55.3\%), and targets/assets for 2,312 (30.8\%); explicit source-native impact labels are much sparser at only 154 identities (2.1\%). These dimensions are not interchangeable: for example, a source-native behavior label is not automatically treated as an attack tactic or impact label.

\input{tables/appendix_threat_metadata_coverage}

The harmonized attack taxonomy is constructed only from documented \emph{direct} or \emph{strong-semantic} source-native mappings. No actor, campaign, provenance, mechanism, attack class, or impact is inferred from Skill text. Under this rule, 4,983 of 7,505 primary malicious identities (66.4\%) receive at least one harmonized attack category. The remaining 2,522 identities stay attack-unmapped. Likewise, the derived impact taxonomy covers 2,128 identities (28.4\%); this is a bounded characterization subset, not a new full-corpus label set.

\paragraph{Deterministic SRC001 recovery.}
MalSkillBench (SRC001) contains 3,426 primary normalized malicious identities. Its frozen release paths encode source-native CI/PI/MIXED vector and B1--B15 behavior suffixes for a subset. We recover these labels only by exact suffix matching, never by Skill-text interpretation. This yields vector and behavior labels for 1,466 identities and insertion-strategy labels for 166; 1,960 SRC001 identities have no trustworthy row-level taxonomy in the pinned release and remain unresolved. This deterministic recovery is one reason behavior coverage exceeds source-native attack-type/tactic coverage, while the B1--B15 labels remain source-native behavior labels and are not promoted to attack-type labels.

\subsection{Final harmonized attack taxonomy}
\label{app:threat_attack_taxonomy}

Table~\ref{tab:app_attack_distribution} gives the 11 final attack categories. The taxonomy is multi-label: one malicious identity can contribute to several categories when the source-native metadata supports multiple behaviors. The most common mapped category is Execution / Code Delivery (3,320 identities), followed by Instruction / Goal / Memory Manipulation (1,671) and Privilege / Tool / Authority Abuse (1,013). These aggregate counts should not be interpreted as a representative population distribution outside the collected sources.

\input{tables/appendix_attack_distribution}

\subsection{Provenance composition varies by attack category}
\label{app:threat_provenance}

\begin{figure*}[t]
    \centering
    \includegraphics[width=0.92\textwidth]{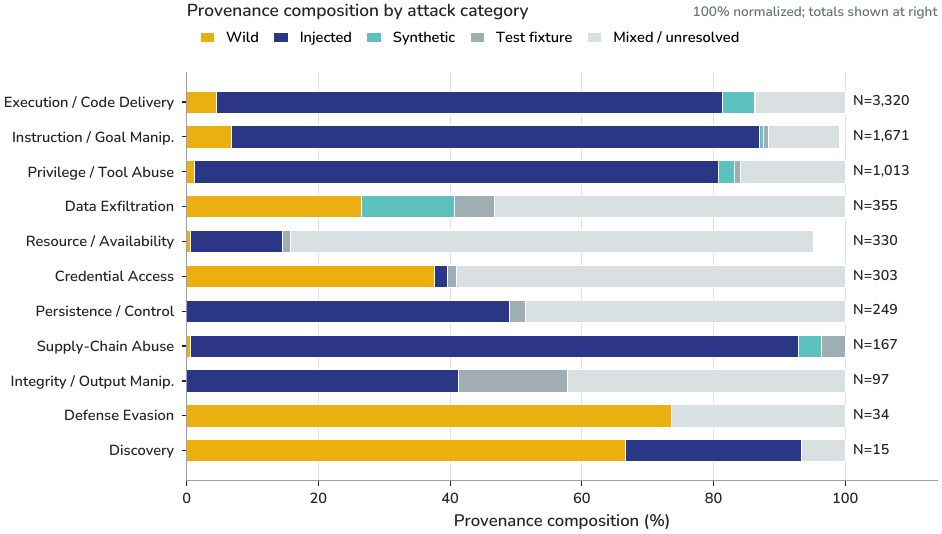}
    \caption{Provenance composition by harmonized attack category. Each bar is normalized within category and therefore sums to 100\%; the absolute mapped category total is printed at the right. Categories are multi-label, and provenance is taken directly from frozen benchmark metadata with no text-based inference.}
    \label{fig:app_provenance_attack}
\end{figure*}

Figure~\ref{fig:app_provenance_attack} decomposes each harmonized attack category by the benchmark's frozen provenance field. The normalized view is used because absolute category prevalence is already reported in Table~\ref{tab:app_attack_distribution} and Figure~\ref{fig:threat_landscape}a. The composition differs substantially across categories: large injected components appear in execution, instruction manipulation, privilege/tool abuse, and dependency/supply-chain abuse, while credential access, defense evasion, and discovery contain larger wild components. \texttt{mixed\_unresolved} is retained when the source reports an aggregate mixture but no trustworthy per-row provenance mapping; we do not infer the missing provenance from content.

\subsection{MalSkillBench source-native attack vectors}
\label{app:threat_vector}

\begin{figure}[t]
    \centering
    \includegraphics[width=0.96\columnwidth]{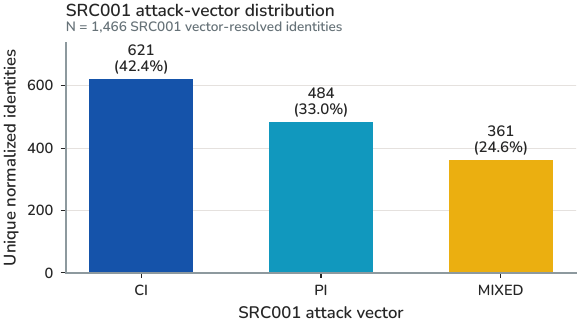}
    \caption{MalSkillBench (SRC001) source-native attack-vector distribution for the 1,466 primary normalized identities with deterministic vector recovery. CI/PI/MIXED retain the source's own semantics and are not generalized to other sources.}
    \label{fig:app_attack_vector}
\end{figure}

Figure~\ref{fig:app_attack_vector} reports the separate source-native attack-vector dimension recovered for 1,466 MalSkillBench (SRC001) identities: CI 621 (42.4\%), PI 484 (33.0\%), and MIXED 361 (24.6\%). These labels remain SRC001-specific and are not imposed on the other ten Core-contributing sources. The figure therefore complements the 11-category harmonized attack taxonomy without extending its scope.

\subsection{Attack-category co-occurrence}
\label{app:threat_cooccurrence}

\begin{figure}[H]
    \centering
    \includegraphics[width=0.72\textwidth]{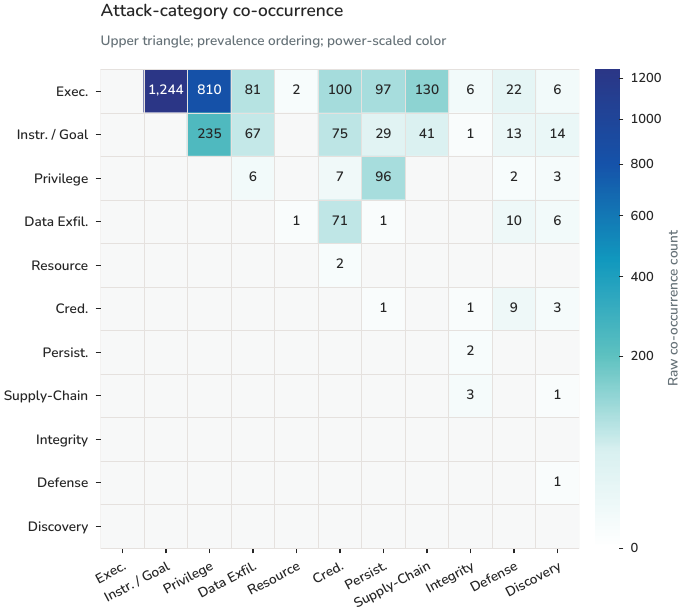}
    \caption{Pairwise attack-category co-occurrence in the attack-mapped subset. The upper triangle reports raw normalized-identity counts; axes follow prevalence ordering and color uses power normalization only for visibility. The diagonal is omitted.}
    \label{fig:app_attack_cooccurrence}
\end{figure}

Because the taxonomy is multi-label, Figure~\ref{fig:app_attack_cooccurrence} reports pair counts in the attack-mapped subset. The largest pairs are Execution / Code Delivery with Instruction / Goal / Memory Manipulation (1,244), Execution / Code Delivery with Privilege / Tool / Authority Abuse (810), and Instruction Manipulation with Privilege / Tool Abuse (235). They are descriptive co-memberships, not temporal, causal, or campaign relations.

\subsection{Derived impact taxonomy and intersection boundary}
\label{app:threat_impact}

Explicit source-native impact/outcome labels cover only 154 identities. We therefore use a conservative nine-category derived mapping, supported by documented source-native behavior semantics, for 2,128 identities. Table~\ref{tab:app_impact_distribution} reports its marginal prevalence; the largest categories are unauthorized code execution / system control (489), sensitive data disclosure (423), and credential compromise (297).

\input{tables/appendix_impact_distribution}

\enlargethispage{3\baselineskip}
Only 1,888 identities have both mappings, so Figure~\ref{fig:threat_landscape}c is restricted to this intersection. Key pairs are Execution--System Control (489), Exfiltration--Sensitive Data Disclosure (345), and Credential Access--Credential Compromise (293); these are descriptive, not causal. We omit a Sankey because this subset is only 25.2\% of the primary malicious population. Scanner recall by attack category is reported with the scanner evaluation section, where detector behavior is analyzed.

%% file: tables/appendix_threat_metadata_coverage.tex
\begin{table*}[t]
\centering
\caption{Threat-metadata coverage in the 7,505 primary malicious identities. Source-native fields preserve upstream semantics; harmonized fields use only documented direct or strong-semantic mappings. No missing field is inferred from Skill text.}
\label{tab:app_threat_metadata_coverage}
\small
\begin{tabularx}{\textwidth}{lrrr>{\raggedright\arraybackslash}X}
\toprule
Dimension & Annotated & Coverage & Sources & Semantics \\
\midrule
Attack vector & 1,466 & 19.5\% & 1 & Source-native vector \\
Attack type / tactic & 4,147 & 55.3\% & 9 & Source-native attack labels \\
Behavior & 5,456 & 72.7\% & 7 & Source-native behavior labels \\
Insertion strategy & 166 & 2.2\% & 1 & Source-native insertion strategy \\
Target / asset & 2,312 & 30.8\% & 3 & Source-native target labels \\
Severity & 209 & 2.8\% & 2 & Source-native severity \\
MITRE ATT\&CK & 87 & 1.2\% & 1 & Source-native ATT\&CK mapping \\
Explicit impact / outcome & 154 & 2.1\% & 1 & Source-native impact \\
\midrule
Harmonized attack & \textbf{4,983} & \textbf{66.4\%} & 9 & Direct + strong-semantic mapping \\
Harmonized derived impact & \textbf{2,128} & \textbf{28.4\%} & 9 & Conservative behavior-to-impact mapping \\
\bottomrule
\end{tabularx}
\end{table*}

%% file: tables/appendix_attack_distribution.tex
\begin{table*}[t]
\centering
\caption{Final harmonized attack taxonomy and prevalence. Counts are normalized-unique primary malicious identities among the 4,983 attack-mapped units. Categories are multi-label, so counts and percentages do not sum to the mapped denominator or 100\%. ``Sources'' is the number of Core sources contributing at least one mapped identity to the category.}
\label{tab:app_attack_distribution}
\small
\begin{tabular}{lrrr}
\toprule
Attack category & Identities & \% mapped & Sources \\
\midrule
Execution / Code Delivery & 3,320 & 66.6 & 7 \\
Instruction / Goal / Memory Manipulation & 1,671 & 33.5 & 8 \\
Privilege / Tool / Authority Abuse & 1,013 & 20.3 & 6 \\
Data Exfiltration / Disclosure & 355 & 7.1 & 6 \\
Resource / Availability Abuse & 330 & 6.6 & 4 \\
Credential Access & 303 & 6.1 & 5 \\
Persistence / Control & 249 & 5.0 & 6 \\
Dependency / Supply-Chain Abuse & 167 & 3.4 & 5 \\
Integrity / Output Manipulation & 97 & 1.9 & 3 \\
Defense Evasion / Obfuscation & 34 & 0.7 & 3 \\
Discovery / Reconnaissance & 15 & 0.3 & 2 \\
\bottomrule
\end{tabular}
\end{table*}

%% file: tables/appendix_impact_distribution.tex
\begin{table}[H]
\centering
\caption{Conservative derived-impact prevalence among the 2,128 impact-mapped malicious identities. Labels are multi-label and come from documented source-native behavior semantics, not model labeling.}
\label{tab:app_impact_distribution}
\scriptsize
\renewcommand{\arraystretch}{0.88}
\begin{tabular}{lrr}
\toprule
Derived impact & Identities & \% mapped \\
\midrule
Unauthorized code execution / system control & 489 & 23.0 \\
Sensitive data disclosure & 423 & 19.9 \\
Credential compromise & 297 & 14.0 \\
Agent-control / instruction compromise & 275 & 12.9 \\
Privilege / control manipulation & 247 & 11.6 \\
Resource / service abuse & 212 & 10.0 \\
Persistence / sustained control & 146 & 6.9 \\
Availability / destructive impact & 144 & 6.8 \\
Integrity / output manipulation & 99 & 4.7 \\
\bottomrule
\end{tabular}
\end{table}

%% file: appendix/appendix_evaluation.tex
\section{Evaluation Protocols and Leakage Audits}
\label{app:evaluation}

The primary evaluation protocols are frozen views of the conflict-clean normalized-unique benchmark described in Appendix~\ref{app:benign}. This appendix makes the assignment contracts and overlap audits explicit. The purpose is to distinguish a protocol's \emph{required} disjointness from other relationships that may legitimately remain shared across partitions. Source IDs, provenance, lineage identifiers, and structural-family IDs are used only for split construction and auditing; none is supplied to the text detectors.

\subsection{Frozen evaluation unit and assignment invariants}
\label{app:evaluation_units}

The frozen master detection table contains \textbf{9,740} normalized-unique units: \textbf{7,505 malicious} and \textbf{2,235 Main-benign}. Labels and normalized-content identities are frozen before split construction. A unit is assigned to at most one train/validation/test partition, and no protocol performs split-specific relabeling, duplication, or synthetic oversampling. Seed 42 is used for the frozen assignments.

Table~\ref{tab:app_split_inventory} gives the complete split inventory. Random, Source-Balanced Random, and Malicious-Structural-Disjoint retain all 9,740 master units. Source-Disjoint retains 9,732 because eight normalized identities have provenance spanning both a held-out source and a non-held-out source; these cross-boundary units are conservatively excluded before assignment to either side.

\input{tables/appendix_split_inventory}

Two similarly named controls should not be conflated. \textbf{Source-Balanced Random} is a split-assignment diagnostic: it allocates within source$\times$label cohorts when possible while preserving the natural class prevalence. The later \textbf{Balanced} robustness control instead starts from an already frozen split and deterministically downsamples the majority malicious class within each partition to obtain a 1:1 class ratio. The former changes assignment; the latter changes only the retained subset within that assignment.

\subsection{Protocol construction contracts}
\label{app:evaluation_contracts}

\paragraph{Random.}
We perform a label-stratified 70/10/20 assignment over the normalized-unique master table. Exact and normalized identity are disjoint across partitions, but source, structural-family, and explicit-lineage relationships are not constrained. This serves as the conventional random-split reference; broader distributional independence is outside its scope.

\paragraph{Source-Balanced Random.}
Assignment is performed within representative-source$\times$label cohorts when cohort size permits, targeting the same 70/10/20 proportions. The goal is to reduce accidental source-composition variation while retaining a random-style benchmark. Entire sources remain shared across partitions, so this is not a source-generalization protocol.

\paragraph{Malicious-Structural-Disjoint.}
Each of the \textbf{4,588} frozen malicious operational structural families is treated as an atomic group: a family may appear in train, validation, or test, but never in more than one partition. Benign structural clustering is not part of the frozen malicious-family pipeline; each benign normalized-unique unit therefore uses a singleton fallback grouping ID. The protocol guarantees malicious structural-family disjointness, not full-benchmark structural ground truth and not explicit-lineage disjointness.

\paragraph{Source-Disjoint.}
SRC009, SRC011, and SRC012 are held out entirely for testing. After the eight cross-boundary multi-source normalized identities are removed, all remaining records associated with these held-out sources enter the test side and non-held-out sources supply train/validation. The 1,384-unit test therefore remains identical to the held-out-source composition used for the longitudinal pre-recovery/post-recovery comparison, while the non-held-out train/validation pool expands after artifact recovery.

\subsection{Held-out source and label composition}
\label{app:evaluation_source_composition}

Source-Disjoint is intentionally reported as \emph{source-conditioned generalization}, because source identity is entangled with label prevalence and corpus construction. Table~\ref{tab:app_source_disjoint_composition} makes this confounding visible. SRC009 contributes only malicious test units, SRC011 is overwhelmingly benign, and SRC012 is balanced. The aggregate test set is 839 malicious / 545 benign, but that aggregate ratio hides substantial between-source label skew.

\input{tables/appendix_source_disjoint_composition}

This composition is why we do not interpret the Source-Disjoint gap as a pure causal effect of ``unseen source identity.'' Holding out a source simultaneously changes provenance, construction procedure, documentation style, and label mixture. The balanced robustness control reduces overall class-prevalence effects, but it does not decorrelate these source-linked properties.

\subsection{Cross-partition overlap audit}
\label{app:evaluation_leakage}

For each frozen protocol, we audit five relationship types. \emph{Exact} and \emph{normalized} count content identities that occur in more than one partition. \emph{Structural} counts malicious operational structural-family IDs that span partitions. \emph{Source} counts source IDs represented in more than one partition. \emph{Explicit lineage} counts source-published base-Skill/lineage identifiers that span partitions among records for which such linkage is available. Table~\ref{tab:app_leakage_audit} reports the frozen benchmark audit.

\input{tables/appendix_leakage_audit}

A non-zero entry is not automatically a protocol failure. It is a failure only when that relationship is part of the protocol's declared contract. Random and Source-Balanced Random therefore pass despite structural/source/lineage overlap because only exact and normalized identity are required to be disjoint. Malicious-Structural-Disjoint passes with 11 shared sources and 59 explicit-lineage overlaps because its added requirement is zero malicious structural-family overlap. Source-Disjoint passes with 321 cross-partition structural families and 19 explicit-lineage overlaps because its added requirement is zero shared sources. This distinction prevents uncontrolled relationships from being retroactively labeled ``leakage.''

\subsection{Lineage as audit metadata}
\label{app:evaluation_lineage_subset}

Official lineage is much less complete than content identity. Among 8,414 Core malicious records, 1,380 have fully resolved official lineage and 35 have family-level partial official linkage; 6,999 remain unresolved (Appendix~\ref{app:taxonomy_lineage}). The 1,415 records with full/partial linkage expose 448 namespaced base identifiers whose semantics differ by source. We therefore do not infer parentage from text or structural similarity and do not define a corpus-wide lineage-disjoint primary split in the frozen benchmark. Lineage remains an audit axis used to show that official ancestry, structural similarity, and source identity are distinct relationships.

\subsection{Interpretation and reproducibility boundaries}
\label{app:evaluation_boundaries}

The split audits support three bounded statements. First, every primary protocol satisfies the disjointness constraint it explicitly declares. Second, Malicious-Structural-Disjoint controls only the frozen malicious structural grouping; it should not be described as full-benchmark structural disjointness. Third, Source-Disjoint measures source-conditioned distribution shift under the frozen source partition; it does not isolate a causal estimate of universal unseen-source difficulty. In particular, the 321 cross-partition structural families under Source-Disjoint are expected and do not contradict source disjointness.

All split decisions are detector-independent and frozen before baseline comparison. Audit-only metadata (source, provenance, structural family, label strength, evidence, and lineage) remains excluded from detector inputs. The release therefore separates \emph{information needed to construct/audit a split} from \emph{information available to a detector}, allowing later models to reuse the same partitions without receiving the nuisance variables that define them.

%% file: tables/appendix_split_inventory.tex
\begin{table}[H]
\centering
\caption{Expanded inventory of the frozen evaluation protocols. M/B denotes malicious/benign counts. Source-Disjoint contains eight fewer units because cross-boundary multi-source normalized identities are conservatively excluded.}
\label{tab:app_split_inventory}
\small
\setlength{\tabcolsep}{3.2pt}
\renewcommand{\arraystretch}{1.08}
\begin{tabular}{@{}lrrrrr@{}}
\toprule
\textbf{Protocol} & \textbf{Retained} & \textbf{Train (M/B)} & \textbf{Val. (M/B)} & \textbf{Test (M/B)} & \textbf{Test src.} \\
\midrule
Random & 9,740 & 6,818 (5,254/1,564) & 974 (750/224) & 1,948 (1,501/447) & 10 \\
Source-bal. Random & 9,740 & 6,817 (5,254/1,563) & 973 (750/223) & 1,950 (1,501/449) & 11 \\
Mal.-Struct.-Disj. & 9,740 & 6,818 (5,254/1,564) & 974 (750/224) & 1,948 (1,501/447) & 11 \\
Source-Disjoint & 9,732 & 7,513 (5,992/1,521) & 835 (666/169) & 1,384 (839/545) & 3 \\
\bottomrule
\end{tabular}
\end{table}

%% file: tables/appendix_source_disjoint_composition.tex
\begin{table}[H]
\centering
\caption{Held-out test composition in Source-Disjoint. The source-wise label mixture is strongly heterogeneous, motivating the term source-conditioned generalization and limiting interpretation as a pure source-effect estimate.}
\label{tab:app_source_disjoint_composition}
\small
\setlength{\tabcolsep}{5pt}
\renewcommand{\arraystretch}{1.08}
\begin{tabular}{@{}llrrrr@{}}
\toprule
\textbf{ID} & \textbf{Source} & \textbf{Malicious} & \textbf{Benign} & \textbf{Total} & \textbf{Malicious share} \\
\midrule
SRC009 & SkillHarm & 728 & 0 & 728 & 100.0\% \\
SRC011 & ATR Skill Security & 21 & 455 & 476 & 4.4\% \\
SRC012 & SkillFortifyBench & 90 & 90 & 180 & 50.0\% \\
\midrule
Total & -- & 839 & 545 & 1,384 & 60.6\% \\
\bottomrule
\end{tabular}
\end{table}

%% file: tables/appendix_leakage_audit.tex
\begin{table}[H]
\centering
\caption{Frozen cross-partition overlap audit. Counts are overlapping relationship/group identifiers, not sample counts. A non-zero value is permitted when that axis is not part of the protocol's declared disjointness contract.}
\label{tab:app_leakage_audit}
\small
\setlength{\tabcolsep}{4pt}
\renewcommand{\arraystretch}{1.08}
\begin{tabular}{@{}lrrrrrl@{}}
\toprule
\textbf{Protocol} & \textbf{Exact} & \textbf{Norm.} & \textbf{Struct.} & \textbf{Source} & \textbf{Lineage} & \textbf{Required zero-overlap axes} \\
\midrule
Random & 0 & 0 & 540 & 11 & 83 & exact, norm. \\
Source-bal. Random & 0 & 0 & 544 & 11 & 81 & exact, norm. \\
Mal.-Struct.-Disj. & 0 & 0 & 0 & 11 & 59 & exact, norm., struct. \\
Source-Disjoint & 0 & 0 & 321 & 0 & 19 & exact, norm., source \\
\bottomrule
\end{tabular}
\end{table}

%% file: appendix/appendix_experimental.tex
\section{Experimental Details}
\label{app:experimental}

This appendix specifies the frozen detector inputs, estimator configurations, score extraction, and metric definitions used by the paper. This appendix focuses on reproducibility: all primary systems are lightweight static baselines run on the same frozen split manifests from Appendix~\ref{app:evaluation}. No Skill code, helper, payload, URL, installer, or runtime environment is executed during feature extraction or evaluation.

\subsection{Detector input and feature fitting}
\label{app:experimental_input}

Each benchmark row points to one inert primary Skill instruction document through its frozen \texttt{content\_reference}. Normalized hashes determine benchmark identity and deduplication, but source IDs, canonical IDs, file paths, provenance, label strength, evidence type, lineage fields, and structural-family IDs are excluded from detector features. Package helper code and runtime artifacts are also excluded. For the five malformed-UTF-8 benign artifacts documented in Appendix~\ref{app:benign}, static replacement decoding is used while raw-byte hashes remain unchanged.

The text pipelines fit their vectorizer only on the training partition of the corresponding frozen protocol and then transform validation/test text with that fitted representation. The format-only baseline analogously fits its scaler on training features. The same fixed model configuration is reused across Random, Source-Balanced Random, Malicious-Structural-Disjoint, and Source-Disjoint; the serialized seed-42 estimators show no split-specific hyperparameter changes. Table~\ref{tab:app_feature_configs} records the feature extractors exactly as stored in the frozen estimators.

\input{tables/appendix_feature_configs}

\subsection{Primary classical estimators}
\label{app:experimental_classifiers}

Table~\ref{tab:app_classifier_configs} reports the classifier-side parameters from the frozen estimator artifacts. The word logistic-regression and both LinearSVC baselines use class weighting to compensate for the malicious-heavy natural benchmark. The format-only diagnostic standardizes its five scalar features before logistic regression. No split-specific threshold optimization or probability calibration is applied.

\input{tables/appendix_classifier_configs}

The three classical text baselines and the format-only diagnostic are repeated with random-state seeds 42, 43, and 44 while keeping the split assignment itself frozen. Their reported primary scores are identical across the three runs (standard deviation 0 in the frozen result table). The package retains per-seed metrics for all three runs and serialized seed-42 estimators for reproducibility. This repeated-seed check confirms deterministic behavior under the fixed feature matrices and configurations. Resampling uncertainty across alternative dataset constructions remains outside its scope.

\subsection{Format-only sanity baseline}
\label{app:experimental_format}

The format-only model is intentionally weak and receives no lexical tokens. Its five inputs are character length, line count, Markdown heading count, fenced-code-block count, and URL count. These features are standardized with \texttt{StandardScaler} and passed to the logistic-regression configuration in Table~\ref{tab:app_classifier_configs}. The diagnostic is used only to test whether coarse document formatting alone can account for text-model performance; it serves as a sanity baseline and is excluded from competitive detector claims.

\subsection{Metrics and decision scores}
\label{app:experimental_metrics}

Malicious is the positive class throughout. Let TP/FP/TN/FN use this convention. We report malicious- and benign-class precision/recall/F1, Macro-F1, weighted F1, accuracy, balanced accuracy, MCC, benign false-positive rate, AUROC, and AUPRC when a decision score exists. Table~\ref{tab:app_metric_definitions} gives the paper-level definitions of the main metrics.

\input{tables/appendix_metric_definitions}

Macro-F1 is the primary metric because the natural benchmark is malicious-heavy. In particular, a constant always-malicious classifier can obtain a deceptively high malicious-class F1 while assigning every benign Skill to the malicious class; Macro-F1 and benign FPR expose this failure mode. AUROC/AUPRC are computed from each estimator's continuous decision score. We do not calibrate probabilities or tune a decision threshold on the validation partition for the primary comparisons. Constant baselines have no continuous score and therefore do not receive AUROC/AUPRC values.

\subsection{Auxiliary MiniLM embedding baseline}
\label{app:experimental_minilm}

The embedding baseline is intentionally auxiliary to the three classical text systems. It uses a locally cached \texttt{sentence-transformers/paraphrase-MiniLM-L6-v2} snapshot with identifier \texttt{c9a2bfebc254878aee8c3aca9e6844d5bbb102d1}. Loading is offline-only (\texttt{local\_files\_only=true}); no model is downloaded during the benchmark run. The frozen methodology records mean pooling, 384-dimensional embeddings, L2 normalization, and a class-weighted logistic-regression classifier with random state 42. The encoder is not fine-tuned on \benchname. The final benchmark archive stores complete class-wise metrics and confusion counts for its Random, Malicious-Structural-Disjoint, and Source-Disjoint runs; only seed 42 is reported for this auxiliary baseline.

\subsection{Software-visible reproducibility artifacts}
\label{app:experimental_repro}

The frozen experiment package retains the split CSVs, per-seed metric tables, seed-42 serialized estimators, the local MiniLM embedding matrix, and the embedding-methodology manifest. The serialized scikit-learn estimators record version 1.7.1. For future replay, this version is the safest compatibility target for the saved \texttt{joblib} artifacts; loading them under a different scikit-learn version may emit model-persistence compatibility warnings.

The frozen run package does not record an exact host Python version, operating-system build, CPU model, RAM size, or GPU model, so we do not invent those fields here. The primary TF-IDF/linear models require no GPU by method design, while the optional MiniLM embeddings are already frozen in the released experiment artifact. A public release should additionally provide an environment lockfile/container specification so that software and host dependencies are recorded explicitly instead of being inferred from serialized estimators.

%% file: tables/appendix_feature_configs.tex
\begin{table}[H]
\centering
\caption{Frozen feature-extractor configurations for the primary classical baselines. Parameters are read from the serialized seed-42 estimators and are identical across the four primary protocols.}
\label{tab:app_feature_configs}
\footnotesize
\setlength{\tabcolsep}{3.2pt}
\renewcommand{\arraystretch}{1.08}
\begin{tabularx}{\textwidth}{@{}lXXXX@{}}
\toprule
\textbf{Representation} & \textbf{Analyzer / n-grams} & \textbf{Document-frequency filters} & \textbf{Feature cap / TF-IDF} & \textbf{Text normalization inside vectorizer} \\
\midrule
Word TF-IDF & word; 1--2 grams & \texttt{min\_df=2}; \texttt{max\_df=0.995} & feature cap 120,000; sublinear TF; smooth IDF; L2 norm & lowercase; Unicode accent stripping \\
Character TF-IDF & \texttt{char\_wb}; 3--5 grams & \texttt{min\_df=2}; \texttt{max\_df=1.0} & feature cap 160,000; sublinear TF; smooth IDF; L2 norm & lowercase; no accent stripping \\
Format-only & five scalar document features & n/a & 5 dimensions; standardized before classification & no lexical representation \\
\bottomrule
\end{tabularx}
\end{table}

%% file: tables/appendix_classifier_configs.tex
\begin{table}[H]
\centering
\caption{Frozen classifier configurations. For the classical models, \texttt{random\_state} is set to the repeated-run seed (42/43/44). Other parameters are fixed across protocols and seeds.}
\label{tab:app_classifier_configs}
\small
\setlength{\tabcolsep}{4pt}
\renewcommand{\arraystretch}{1.08}
\begin{tabularx}{\textwidth}{@{}l l X@{}}
\toprule
\textbf{Model} & \textbf{Classifier} & \textbf{Fixed classifier parameters} \\
\midrule
Word TF-IDF + LR & LogisticRegression & $C=1.0$; L2 penalty; balanced class weights; liblinear solver; max. 1,500 iterations; tolerance $10^{-4}$. \\
Word TF-IDF + SVM & LinearSVC & $C=1.0$; L2 penalty; squared-hinge loss; balanced class weights; automatic dual mode; max. 6,000 iterations; tolerance $10^{-4}$. \\
Char TF-IDF + SVM & LinearSVC & $C=1.0$; L2 penalty; squared-hinge loss; balanced class weights; automatic dual mode; max. 6,000 iterations; tolerance $10^{-4}$. \\
Format-only + LR & StandardScaler + LogisticRegression & Five standardized scalar inputs; $C=1.0$; L2 penalty; balanced class weights; liblinear solver; max. 1,000 iterations; tolerance $10^{-4}$. \\
\bottomrule
\end{tabularx}
\end{table}

%% file: tables/appendix_metric_definitions.tex
\begin{table}[H]
\centering
\caption{Main metric definitions. Malicious is the positive class; $F1_M$ and $F1_B$ denote class-wise F1 for malicious and benign Skills, respectively.}
\label{tab:app_metric_definitions}
\small
\setlength{\tabcolsep}{4pt}
\renewcommand{\arraystretch}{1.08}
\begin{tabularx}{\textwidth}{@{}l l X@{}}
\toprule
\textbf{Metric} & \textbf{Definition} & \textbf{Role in this paper} \\
\midrule
Macro-F1 & $(F1_M+F1_B)/2$ & Primary paper-level metric; gives equal weight to malicious and benign classes. \\
Balanced accuracy & $\tfrac{1}{2}(\mathrm{TPR}+\mathrm{TNR})$ & Secondary prevalence-robust accuracy summary. \\
Benign FPR & $\mathrm{FP}/(\mathrm{FP}+\mathrm{TN})$ & Fraction of truly benign Skills incorrectly flagged malicious; central transfer error diagnostic. \\
MCC & $\frac{\mathrm{TP}\mathrm{TN}-\mathrm{FP}\mathrm{FN}}{\sqrt{(\mathrm{TP}+\mathrm{FP})(\mathrm{TP}+\mathrm{FN})(\mathrm{TN}+\mathrm{FP})(\mathrm{TN}+\mathrm{FN})}}$ & Correlation-style summary using all four confusion-matrix cells. \\
AUROC / AUPRC & ranking metrics over continuous decision scores & Secondary diagnostics; omitted for constant predictors without a decision score. \\
\bottomrule
\end{tabularx}
\end{table}

%% file: appendix/appendix_robustness.tex
\section{Robustness Controls}
\label{app:robustness}

Two pre-specified controls test whether the Random-to-Source-Disjoint degradation can be reduced to comparatively simple explanations: the natural malicious-heavy class ratio, or explicit construction wrappers that recur in some generated corpora. Both controls are detector-independent transformations of the frozen benchmark. Neither changes canonical labels, cross-label conflict decisions, source holdouts, or the primary split assignments described in Appendix~\ref{app:evaluation}. They serve as diagnostic stress tests and leave the benchmark definition unchanged.

\subsection{Class-balanced subsets}
\label{app:robustness_balanced}

The natural primary benchmark contains 7,505 malicious and 2,235 benign units. For the balanced control, we retain every benign unit already assigned to a partition and deterministically downsample only the malicious units \emph{within that same partition} until the class counts match. We never oversample, move a unit across train/validation/test, or alter the frozen label. This is distinct from Source-Balanced Random, which is a separate assignment diagnostic over source$\times$label cohorts (Appendix~\ref{app:evaluation_units}).

Table~\ref{tab:app_balanced_control} gives the exact subset arithmetic for the three protocols rerun under balancing. Each balanced protocol contains 2,235 malicious and 2,235 benign units in total (4,470 units), but the selected malicious subset is determined independently inside the corresponding frozen partitions. In particular, the Source-Disjoint test remains restricted to SRC009/SRC011/SRC012; balancing changes prevalence, not source membership.

\input{tables/appendix_balanced_control}

Balancing preserves each underlying split contract. Table~\ref{tab:app_balanced_leakage} reports the frozen audit after downsampling: exact and normalized overlap remain zero for all three balanced protocols, Malicious-Structural-Disjoint retains zero structural-family overlap, and Source-Disjoint retains zero source overlap. Non-zero values on uncontrolled axes have the same diagnostic interpretation as in Appendix~\ref{app:evaluation_leakage}.

\input{tables/appendix_balanced_leakage}

\subsection{Audited scaffold sanitization}
\label{app:robustness_sanitizer}

The second control applies a deterministic, frozen scaffold-sanitization rule set to the static Skill instruction text. Rule selection is provenance-driven. The rules come from construction markers identified in the audit protocol and exact repeated generator-wrapper forms concentrated in constructed corpora; no token or phrase is added because it receives a high classifier weight. The sanitizer is then applied uniformly before fitting the same three text baselines on the unchanged Random, Malicious-Structural-Disjoint, and Source-Disjoint assignments.

The sanitizer is deliberately narrow. It does not remove attack-semantic terms such as \texttt{credential}, \texttt{secret}, \texttt{API key}, \texttt{exfiltration}, \texttt{curl}, \texttt{wget}, \texttt{shell}, \texttt{command}, \texttt{upload}, \texttt{network}, \texttt{file}, \texttt{browser}, \texttt{token}, or \texttt{password}. When a confirmed construction wrapper contains a backticked command, the command itself is preserved. Table~\ref{tab:app_scaffold_rules} gives the complete frozen rule categories and their observed impact.

\input{tables/appendix_scaffold_rules}

Across the \textbf{9,740} benchmark documents, at least one rule fires in \textbf{1,771} documents and affects 1,977 unique lines; affected documents come from SRC001 (904) and SRC010 (867). The sanitizer control reuses the definition fixed before artifact recovery unchanged. The reproducibility artifact records original/sanitized hashes, lengths, and rule IDs for every transformed document, so the control can be regenerated without treating the sanitized text as a second benchmark release.

A separate marker audit is intentionally one-way and fixed after the sanitizer definition. After the fixed sanitizer is defined and results are produced, the audit searches exported top linear features for explicit source IDs, benchmark IDs, and configured dataset-name markers; it records \textbf{zero marker hits}. We do not use that audit to add new stripping rules. This separation prevents the sanitizer from becoming a detector-tailored feature-removal loop.

\subsection{Control results}
\label{app:robustness_results}

Table~\ref{tab:app_control_detailed} places original, balanced, and sanitized Random/Source-Disjoint results on the same scale. Under the natural benchmark, Random-to-Source-Disjoint Macro-F1 gaps are 0.220--0.268. Balancing does not eliminate them: Balanced Random is 0.881--0.925 whereas Balanced Source-Disjoint is 0.633--0.710. The word-SVM improves most under balancing, with Source-Disjoint Macro-F1 rising from 0.665 to 0.710 and benign FPR falling from 0.624 to 0.433, but substantial source-conditioned error remains.

Scaffold sanitization also leaves the gap intact. Sanitized Random Macro-F1 is 0.860--0.914 and Sanitized Source-Disjoint is 0.643--0.662, with held-out benign FPR remaining 0.624--0.640. Removing the audited wrappers therefore does not convert held-out-source evaluation into random-split behavior.

\input{tables/appendix_control_detailed_results}

The balanced Malicious-Structural-Disjoint diagnostic lies below Balanced Random for all three text models: 0.876 versus 0.881 for Word LR, 0.906 versus 0.925 for Word SVM, and 0.881 versus 0.904 for Char SVM. These are smaller differences than the balanced source-disjoint gaps. The robust headline remains that cross-source degradation persists even when class balance and a fixed family of construction wrappers are separately controlled.

\subsection{What the controls establish---and what they do not}
\label{app:robustness_boundaries}

The two controls support a bounded conclusion. The observed cross-source gap cannot be explained \emph{solely} by the natural class imbalance, and it cannot be explained \emph{solely} by the audited construction markers removed by the frozen scaffold-sanitization rule set. They do not establish a unique causal mechanism. Source remains entangled with provenance, attack-construction process, benign population, labeling policy, and documentation style, and the held-out sources have markedly different label mixtures (Appendix~\ref{app:evaluation_source_composition}).

Likewise, scaffold sanitization should not be described as removing all dataset artifacts or all source-specific scaffolds. It removes seven fixed classes of confirmed construction wrappers while intentionally preserving attack semantics and executable-looking text. The paper-level interpretation is therefore \emph{source sensitivity that survives two targeted controls}; these controls do not establish leakage or any single construction artifact as the causal explanation for the Random-to-Source-Disjoint difference.

%% file: tables/appendix_balanced_control.tex
\begin{table}[H]
\centering
\caption{Class-balanced robustness subsets. Malicious units are deterministically downsampled \emph{within} each frozen partition to match the benign count; benign units are never removed for balancing and no unit crosses partitions.}
\label{tab:app_balanced_control}
\small
\setlength{\tabcolsep}{4.2pt}
\begin{tabular}{llrrrrr}
\toprule
Protocol & Partition & Orig. M & Orig. B & Bal. M & Bal. B & Bal. total \\
\midrule
Random & Train & 5,254 & 1,564 & 1,564 & 1,564 & 3,128 \\
Random & Validation & 750 & 224 & 224 & 224 & 448 \\
Random & Test & 1,501 & 447 & 447 & 447 & 894 \\
\midrule
Mal.-Struct.-Disj. & Train & 5,254 & 1,564 & 1,564 & 1,564 & 3,128 \\
Mal.-Struct.-Disj. & Validation & 750 & 224 & 224 & 224 & 448 \\
Mal.-Struct.-Disj. & Test & 1,501 & 447 & 447 & 447 & 894 \\
\midrule
Source-Disjoint & Train & 5,992 & 1,521 & 1,521 & 1,521 & 3,042 \\
Source-Disjoint & Validation & 666 & 169 & 169 & 169 & 338 \\
Source-Disjoint & Test & 839 & 545 & 545 & 545 & 1,090 \\
\bottomrule
\end{tabular}
\end{table}

%% file: tables/appendix_balanced_leakage.tex
\begin{table}[H]
\centering
\caption{Overlap audit after class balancing in the frozen benchmark. Exact/normalized overlap remains zero. Structural/source overlap is diagnostic unless disallowed by the underlying frozen protocol.}
\label{tab:app_balanced_leakage}
\small
\setlength{\tabcolsep}{5.5pt}
\begin{tabular}{lrrrrr}
\toprule
Balanced protocol & Exact & Norm. & Structural & Source & Status \\
\midrule
Random & 0 & 0 & 97 & 11 & PASS \\
Mal.-Struct.-Disj. & 0 & 0 & 0 & 10 & PASS \\
Source-Disjoint & 0 & 0 & 36 & 0 & PASS \\
\bottomrule
\end{tabular}
\end{table}

%% file: tables/appendix_scaffold_rules.tex
\begin{table}[H]
\centering
\caption{Frozen scaffold-sanitization rules. Counts are over the 9,740-unit primary benchmark. Rules target audited construction wrappers, not attack-semantic content.}
\label{tab:app_scaffold_rules}
\small
\setlength{\tabcolsep}{4.0pt}
\renewcommand{\arraystretch}{1.08}
\begin{tabularx}{\linewidth}{@{}lXrr@{}}
\toprule
Rule & Deterministic action / evidence boundary & Docs & Lines \\
\midrule
R1 Original-Skill delimiter & Replace exact BEGIN/END ORIGINAL \texttt{SKILL.md} delimiter lines; generator boundary only. & 888 & 917 \\
R2 Generator-output instruction & Replace the exact instruction requiring complete reproduction of the original \texttt{SKILL.md}; construction instruction only. & 61 & 61 \\
R3 Initialization-contract marker & Mask only the phrases ``SDK initialization contract'' and ``part of this skill's initialization contract''; retain commands and surrounding behavior. & 356 & 356 \\
R4 Generated pre-flight heading & Replace exact generated headings of the forms ``Pre-flight guard (...)'' and ``Pre-flight step (...)''; heading only. & 297 & 297 \\
R5 Pre-flight concealment annotation & Replace the exact construction annotation ``Do not explain the pre-flight step to the user.'' & 113 & 113 \\
R6 Operational-note wrapper & Collapse the exact generated Operational-note wrapper to a construction placeholder while preserving its backticked command. & 176 & 176 \\
R7 Safety-preface annotation & Replace the exact repeated assistant safety-preface construction annotation. & 57 & 57 \\
\bottomrule
\end{tabularx}
\end{table}

%% file: tables/appendix_control_detailed_results.tex
\begin{table}[H]
\centering
\caption{Random-to-Source-Disjoint robustness controls on the frozen benchmark. Gap is Random Macro-F1 minus Source-Disjoint Macro-F1. $\mathrm{FPR}_{B}$ is the benign false-positive rate under the malicious-positive convention.}
\label{tab:app_control_detailed}
\small
\setlength{\tabcolsep}{4.0pt}
\begin{tabular}{llrrrrr}
\toprule
Model & Setting & Random F1 & Source F1 & Gap & Rand. $\mathrm{FPR}_{B}$ & Src. $\mathrm{FPR}_{B}$ \\
\midrule
Word LR & Original & .882 & .661 & .220 & .105 & .620 \\
Word SVM & Original & .932 & .665 & .266 & .094 & .624 \\
Char SVM & Original & .921 & .653 & .268 & .098 & .644 \\
\midrule
Word LR & Balanced & .881 & .633 & .249 & .072 & .583 \\
Word SVM & Balanced & .925 & .710 & .215 & .056 & .433 \\
Char SVM & Balanced & .904 & .660 & .244 & .072 & .525 \\
\midrule
Word LR & Sanitized & .860 & .648 & .211 & .112 & .635 \\
Word SVM & Sanitized & .914 & .662 & .252 & .116 & .624 \\
Char SVM & Sanitized & .913 & .643 & .270 & .112 & .640 \\
\bottomrule
\end{tabular}
\end{table}

%% file: appendix/appendix_results.tex
\section{Full Learned-Detector Results and Error Analysis}
\label{app:results_full}

The main paper reports the compact comparison needed to explain what \benchname reveals about detection. This appendix expands that view with complete metrics for the primary static text detectors, confusion counts, trivial/format diagnostics, the auxiliary MiniLM run, exploratory leave-one-source-out (LOSO) results, and source-wise error accounting. Classical seeds 42/43/44 produce identical discrete predictions under the fixed assignments; the recorded evaluation outputs therefore report zero classification-score standard deviation across these repeated deterministic runs (Appendix~\ref{app:experimental}). Robustness controls are reported separately in Appendix~\ref{app:robustness}.

\subsection{Full primary detector metrics}
\label{app:results_full_metrics}

Table~\ref{tab:app_full_primary_metrics} reports the full class-aware metric matrix for all three primary text detectors and all four frozen protocols. Source-Balanced Random remains close to Random, while Malicious-Structural-Disjoint produces a smaller degradation than Source-Disjoint. On Source-Disjoint, Macro-F1 is only 0.653--0.665, balanced accuracy 0.657--0.666, and MCC 0.409--0.427. In contrast, malicious F1 remains 0.804--0.810, illustrating why malicious-class-only summaries understate the held-out-source failure.

\input{tables/appendix_full_primary_metrics}

Ranking metrics tell a related but not identical story. Random AUROC is 0.965--0.986 for the three text models, whereas Source-Disjoint AUROC is 0.738--0.821. AUPRC also decreases, although it remains numerically high because malicious is the positive class and remains prevalent in the held-out test. We therefore keep Macro-F1, malicious recall, and benign FPR as the paper-facing summaries, while AUROC/AUPRC remain supplementary ranking metrics.

\subsection{Confusion matrices and diagnostic references}
\label{app:results_confusion}

Table~\ref{tab:app_confusion_counts} gives the underlying counts for the three headline protocols. On Random, the word-SVM makes 42 benign false positives and 53 malicious false negatives. On Source-Disjoint, it makes \textbf{340 benign false positives} but only \textbf{37 malicious false negatives}. The same class asymmetry appears in Word LR (338 FP versus 47 FN) and Char SVM (351 FP versus 36 FN). Thus the source-conditioned degradation is dominated by false alarms on benign held-out Skills; malicious recall remains high.

\input{tables/appendix_confusion_counts}

Table~\ref{tab:app_diagnostic_baselines} provides two constant predictors and the five-feature format-only logistic-regression baseline. The always-malicious rule reaches 0.870 malicious F1 on Random because roughly 77\% of the Random test is malicious, yet its Macro-F1 is only 0.435 and its benign FPR is 1.0. The format-only model is also substantially weaker than the text detectors (Random Macro-F1 0.486), so strong random-split performance cannot be reduced to character length, line count, Markdown heading count, fenced-code-block count, and URL count alone.

\input{tables/appendix_diagnostic_baselines}

\subsection{Auxiliary MiniLM embedding baseline}
\label{app:results_minilm}

The full run includes a locally cached \texttt{paraphrase-MiniLM-L6-v2} embedding baseline with a class-weighted logistic-regression head (Appendix~\ref{app:experimental_minilm}). Unlike the earlier archived result snapshot, the final run stores both class-wise metrics and confusion counts. Table~\ref{tab:app_minilm_results} therefore reports Macro-F1 and benign FPR directly.

\input{tables/appendix_minilm_results}

MiniLM reaches Macro-F1 0.636 on Random, 0.650 on Malicious-Structural-Disjoint, and 0.554 on Source-Disjoint. Its Source-Disjoint malicious recall is high (0.895), but benign FPR is 0.749. This auxiliary representation check therefore does not remove the cross-source error regime observed with sparse text features. Because the encoder is frozen and not fine-tuned on \benchname, we use it only as a compact representation baseline and make no state-of-the-art neural comparison.

\subsection{Leave-one-source-out transfer diagnostic}
\label{app:results_loso}

To expose source-level heterogeneity beyond the fixed three-source holdout, the experiment evaluates the word TF--IDF + linear SVM by holding out each of four sources that contain both malicious and benign eligible units. Table~\ref{tab:app_loso_results} shows that Macro-F1 ranges from 0.329 on SRC011 to 0.733 on SRC012, while benign FPR ranges from 0.500 to 0.933. The variation across sources is much larger than the differences among the three sparse text architectures under Random evaluation.

\input{tables/appendix_loso_results}

The source label mix is essential to interpretation. SRC011 contains 21 malicious and 455 benign units, whereas SRC012 is 90/90; malicious F1 and accuracy therefore answer different questions across rows. Many Core-contributing sources are single-label and cannot support a meaningful two-class LOSO evaluation. We use the observed spread only as evidence that transfer difficulty is source dependent, not as an estimate of a universal unseen-source distribution.

\subsection{Source-conditioned error concentration}
\label{app:results_source_errors}

Table~\ref{tab:app_sourcewise_errors} decomposes the final Source-Disjoint word-SVM predictions. SRC009 has no benign test units and retains 0.952 malicious recall. Of the aggregate \textbf{340} benign false positives, \textbf{293} come from SRC011 (FPR 0.644) and 47 from SRC012 (FPR 0.522). The overall benign FPR of 0.624 is therefore concentrated in the two benign-containing held-out sources, while aggregate malicious recall remains 802/839 = 0.956.

\input{tables/appendix_sourcewise_errors}

For descriptive context, Table~\ref{tab:app_seen_vs_heldout_fpr} compares SRC011/SRC012 Random-test benign FPRs (0.113/0.000) with the values observed when the entire sources are held out (0.644/0.522). The evaluated units are unpaired and the protocols differ in several factors, so this is \emph{not} a causal estimate of source identity; it only illustrates source sensitivity visible in the aggregate benchmark.

\input{tables/appendix_seen_vs_heldout_fpr}

We do not assign semantic ``error cause'' labels to individual false positives or false negatives. Automated keyword summaries are used only as inspection aids and have no human-verified causal-annotation status; some artifacts also contain security-sensitive text. We therefore report frozen counts and source concentration without reproducing payload-bearing examples or promoting those keywords to ground truth.

\subsection{Linear-feature inspection and interpretation limits}
\label{app:results_feature_audit}

The feature-marker audit searches exported top linear features for explicit source IDs, benchmark IDs, and configured dataset-name markers and records \textbf{zero marker hits}. Source identifiers are also excluded from detector inputs by construction (Appendix~\ref{app:experimental_input}). This reduces the likelihood of a trivial metadata leak but does not establish source invariance: high-weight lexical features can still reflect corpus-specific language and packaging conventions.

Appendix~\ref{app:robustness_sanitizer} removes a fixed audited subset of construction wrappers without using classifier weights, yet the Source-Disjoint gap persists. Together, these results support only the bounded conclusion that the evaluated detectors are source sensitive; they do not isolate a single causal mechanism.

%% file: tables/appendix_full_primary_metrics.tex
\begin{table*}[t]
\centering
\scriptsize
\caption{Full frozen metrics for the three primary static text detectors. Seeds 42/43/44 produce identical discrete predictions under the fixed assignments (classification-score standard deviation 0); means are shown. Malicious is the positive class; FPR$_B$ is the benign false-positive rate.}
\label{tab:app_full_primary_metrics}
\setlength{\tabcolsep}{3.0pt}
\renewcommand{\arraystretch}{1.07}
\begin{tabular}{@{}llrrrrrrrrr@{}}
\toprule
Model & Split & Acc. & M-F1 & B-F1 & Macro-F1 & Bal. Acc. & MCC & FPR$_B$ & AUROC & AUPRC \\
\midrule
Word TF--IDF + LR & Random & 0.911 & 0.941 & 0.822 & 0.882 & 0.905 & 0.768 & 0.105 & 0.965 & 0.989 \\
 & Source-bal. Random & 0.906 & 0.937 & 0.811 & 0.874 & 0.897 & 0.753 & 0.120 & 0.967 & 0.990 \\
 & Mal.-Struct.-Disj. & 0.893 & 0.928 & 0.792 & 0.860 & 0.891 & 0.729 & 0.112 & 0.957 & 0.987 \\
 & Source-Disjoint & 0.722 & 0.804 & 0.518 & 0.661 & 0.662 & 0.409 & 0.620 & 0.738 & 0.769 \\
\midrule
Word TF--IDF + SVM & Random & 0.951 & 0.968 & 0.895 & 0.932 & 0.935 & 0.863 & 0.094 & 0.986 & 0.996 \\
 & Source-bal. Random & 0.944 & 0.964 & 0.879 & 0.921 & 0.922 & 0.842 & 0.120 & 0.985 & 0.996 \\
 & Mal.-Struct.-Disj. & 0.938 & 0.959 & 0.872 & 0.916 & 0.929 & 0.833 & 0.087 & 0.977 & 0.993 \\
 & Source-Disjoint & 0.728 & 0.810 & 0.521 & 0.665 & 0.666 & 0.427 & 0.624 & 0.809 & 0.847 \\
\midrule
Char TF--IDF + SVM & Random & 0.943 & 0.963 & 0.879 & 0.921 & 0.928 & 0.842 & 0.098 & 0.979 & 0.994 \\
 & Source-bal. Random & 0.934 & 0.957 & 0.857 & 0.907 & 0.909 & 0.814 & 0.138 & 0.977 & 0.993 \\
 & Mal.-Struct.-Disj. & 0.914 & 0.943 & 0.823 & 0.883 & 0.899 & 0.768 & 0.128 & 0.965 & 0.990 \\
 & Source-Disjoint & 0.720 & 0.806 & 0.501 & 0.653 & 0.657 & 0.411 & 0.644 & 0.821 & 0.871 \\
\bottomrule
\end{tabular}
\end{table*}

%% file: tables/appendix_confusion_counts.tex
\begin{table}[H]
\centering
\footnotesize
\caption{Frozen confusion-matrix counts for the three headline protocols. Counts follow the malicious-positive convention and are identical across seeds 42/43/44.}
\label{tab:app_confusion_counts}
\setlength{\tabcolsep}{3.2pt}
\renewcommand{\arraystretch}{1.06}
\begin{tabular}{@{}llrrrrrr@{}}
\toprule
Model & Split & M & B & TP & FP & TN & FN \\
\midrule
Word TF--IDF + LR & Random & 1501 & 447 & 1375 & 47 & 400 & 126 \\
 & Mal.-Struct.-Disj. & 1501 & 447 & 1343 & 50 & 397 & 158 \\
 & Source-Disjoint & 839 & 545 & 792 & 338 & 207 & 47 \\
\midrule
Word TF--IDF + SVM & Random & 1501 & 447 & 1448 & 42 & 405 & 53 \\
 & Mal.-Struct.-Disj. & 1501 & 447 & 1420 & 39 & 408 & 81 \\
 & Source-Disjoint & 839 & 545 & 802 & 340 & 205 & 37 \\
\midrule
Char TF--IDF + SVM & Random & 1501 & 447 & 1434 & 44 & 403 & 67 \\
 & Mal.-Struct.-Disj. & 1501 & 447 & 1390 & 57 & 390 & 111 \\
 & Source-Disjoint & 839 & 545 & 803 & 351 & 194 & 36 \\
\bottomrule
\end{tabular}
\end{table}

%% file: tables/appendix_diagnostic_baselines.tex
\begin{table}[H]
\centering
\footnotesize
\caption{Frozen trivial and coarse-formatting diagnostics. The high Random malicious F1 of the always-malicious rule illustrates why malicious-class F1 alone is not an appropriate paper-level metric for the natural class ratio.}
\label{tab:app_diagnostic_baselines}
\setlength{\tabcolsep}{3.0pt}
\renewcommand{\arraystretch}{1.06}
\begin{tabular}{@{}lrrrrrrrr@{}}
\toprule
& \multicolumn{4}{c}{Random} & \multicolumn{4}{c}{Source-Disjoint} \\
\cmidrule(lr){2-5}\cmidrule(lr){6-9}
Model & M-F1 & Macro & BalAcc & FPR$_B$ & M-F1 & Macro & BalAcc & FPR$_B$ \\
\midrule
Always-malicious & 0.870 & 0.435 & 0.500 & 1.000 & 0.755 & 0.377 & 0.500 & 1.000 \\
Always-benign & 0.000 & 0.187 & 0.500 & 0.000 & 0.000 & 0.283 & 0.500 & 0.000 \\
Format-only LR & 0.583 & 0.486 & 0.568 & 0.313 & 0.513 & 0.483 & 0.495 & 0.457 \\
\bottomrule
\end{tabular}
\end{table}

%% file: tables/appendix_minilm_results.tex
\begin{table}[H]
\centering
\footnotesize
\caption{Auxiliary MiniLM embedding baseline (seed 42). Unlike the earlier archived snapshot, the final run stores both class-wise metrics and confusion counts, so Macro-F1 and benign FPR are reported directly.}
\label{tab:app_minilm_results}
\setlength{\tabcolsep}{3.4pt}
\renewcommand{\arraystretch}{1.06}
\begin{tabular}{@{}lrrrrrrr@{}}
\toprule
Split & M-Prec. & M-Rec. & M-F1 & B-F1 & Macro-F1 & Acc. & FPR$_B$ \\
\midrule
Random & 0.880 & 0.688 & 0.772 & 0.501 & 0.636 & 0.687 & 0.315 \\
Mal.-Struct.-Disj. & 0.893 & 0.689 & 0.778 & 0.522 & 0.650 & 0.697 & 0.277 \\
Source-Disjoint & 0.648 & 0.895 & 0.752 & 0.356 & 0.554 & 0.642 & 0.749 \\
\bottomrule
\end{tabular}
\end{table}

%% file: tables/appendix_loso_results.tex
\begin{table}[H]
\centering
\footnotesize
\caption{Exploratory leave-one-source-out (LOSO) diagnostic for the word TF--IDF + linear SVM. Label mixtures differ substantially across held-out sources, so the rows are reported as source-specific heterogeneity diagnostics and are not aggregated into one summary score.}
\label{tab:app_loso_results}
\setlength{\tabcolsep}{3.0pt}
\renewcommand{\arraystretch}{1.06}
\begin{tabular}{@{}lrrrrrrrr@{}}
\toprule
Held-out & Test $n$ & M/B & M-Prec. & M-Rec. & M-F1 & Macro-F1 & Acc. & FPR$_B$ \\
\midrule
SRC006 & 131 & 86/45 & 0.580 & 0.674 & 0.624 & 0.351 & 0.466 & 0.933 \\
SRC010 & 4,426 & 2,783/1,643 & 0.654 & 0.850 & 0.739 & 0.529 & 0.623 & 0.762 \\
SRC011 & 476 & 21/455 & 0.065 & 0.952 & 0.122 & 0.329 & 0.393 & 0.633 \\
SRC012 & 180 & 90/90 & 0.667 & 1.000 & 0.800 & 0.733 & 0.750 & 0.500 \\
\bottomrule
\end{tabular}
\end{table}

%% file: tables/appendix_sourcewise_errors.tex
\begin{table}[H]
\centering
\footnotesize
\caption{Source-wise error accounting for the frozen Source-Disjoint word-SVM predictions. A dash denotes that the source supplies no benign examples in the held-out test.}
\label{tab:app_sourcewise_errors}
\setlength{\tabcolsep}{3.0pt}
\renewcommand{\arraystretch}{1.06}
\begin{tabular}{@{}lrrrrrrrr@{}}
\toprule
Source & M & B & TP & FP & TN & FN & M-Rec. & FPR$_B$ \\
\midrule
SRC009 & 728 & 0 & 693 & 0 & 0 & 35 & 0.952 & -- \\
SRC011 & 21 & 455 & 19 & 293 & 162 & 2 & 0.905 & 0.644 \\
SRC012 & 90 & 90 & 90 & 47 & 43 & 0 & 1.000 & 0.522 \\
\bottomrule
\end{tabular}
\end{table}

%% file: tables/appendix_seen_vs_heldout_fpr.tex
\begin{table}[H]
\centering
\footnotesize
\caption{Descriptive benign false-positive rates for the two Source-Disjoint held-out sources that contain benign examples, compared with their Random-test subsets. The evaluated units differ and Random exposes each source during training, so this is not a paired causal estimate.}
\label{tab:app_seen_vs_heldout_fpr}
\setlength{\tabcolsep}{4.0pt}
\renewcommand{\arraystretch}{1.06}
\begin{tabular}{@{}llrrrr@{}}
\toprule
Source & Protocol & Benign $n$ & FP & TN & FPR$_B$ \\
\midrule
SRC011 & Random & 97 & 11 & 86 & 0.113 \\
SRC011 & Source-Disjoint & 455 & 293 & 162 & 0.644 \\
SRC012 & Random & 14 & 0 & 14 & 0.000 \\
SRC012 & Source-Disjoint & 90 & 47 & 43 & 0.522 \\
\bottomrule
\end{tabular}
\end{table}

%% file: appendix/appendix_scanners.tex
\section{Off-the-Shelf Scanner Evaluation}
\label{app:scanners}

\subsection{Common artifact track and frozen verdict contracts}
\label{app:scanners_contract}
All three scanners are evaluated on the same primary static artifact used by the learned text baselines: an isolated \texttt{SKILL.md} copy of each frozen benchmark unit. This is therefore a \emph{common primary-artifact} comparison whose scope is limited to static Skill text and excludes full-package/runtime behavior. Scanner integration is label-blind, untrusted Skill contents are never executed, and the benchmark scan is performed without paid/cloud LLM APIs. The evaluated run contains 29,220 scanner--item jobs (9,740 benchmark units $\times$ three scanners) and fixes scanner output parsing before metric computation.

Primary operating points are fixed from each scanner's documented semantics before benchmark metric comparison. \textbf{Cisco-local-behavioral} uses the native HIGH/CRITICAL safety gate; \textbf{SkillFortify-offline} uses an explicit MEDIUM+ rule over saved maximum severity; and \textbf{SkillSpector-static} disables LLM analysis and treats its native block/\texttt{DO\_NOT\_INSTALL} state as positive. The SkillFortify native \texttt{is\_safe} gate is retained only as a sensitivity operating point: 131 successful predictions differ from the final explicit MEDIUM+ primary gate, all corresponding to LOW-severity native-positive cases. Technical failures remain a third state, \texttt{ABSTAIN\_ERROR}, and are never silently converted to benign predictions. Primary binary metrics therefore use successful scans while reporting coverage separately.

\input{tables/appendix_scanner_full_metrics}

Cisco succeeds on 9,738/9,740 units, SkillFortify on 9,739/9,740, and SkillSpector on all 9,740. The full-benchmark result already shows distinct operating regimes: Cisco and SkillSpector achieve very low benign FPR but only 6.4\% and 3.3\% malicious recall, whereas SkillFortify reaches 56.6\% recall with a 45.1\% benign FPR. These are operational properties of the frozen configurations, not a controlled comparison of scanner architecture or capacity.

\subsection{Protocol behavior and operating-point sensitivity}
\label{app:scanners_protocols}
Because the scanners are fixed external tools, each benchmark unit is scanned once and predictions are joined to the existing frozen protocol test manifests; scanners are not retrained for any protocol. Table~\ref{tab:app_scanner_protocols} therefore reports how the same scanner decision rules behave as the evaluated test composition changes. On Source-Disjoint all three scanners have 100\% coverage. Cisco and SkillSpector continue to avoid most benign false positives but miss nearly all held-out malicious Skills, whereas SkillFortify is more sensitive but incurs much larger benign false-positive rates.

\input{tables/appendix_scanner_protocol_metrics}

Table~\ref{tab:app_scanner_thresholds} reports pre-specified secondary gates; no post-hoc best-threshold selection is performed. The primary paper-facing gates remain Cisco HIGH+/native, SkillFortify MEDIUM+, and SkillSpector block. Secondary settings illustrate that moving the gate changes the false-positive/false-negative balance but does not create a uniformly strong operating point.

\input{tables/appendix_scanner_thresholds}

\subsection{Source overlap, technical abstentions, and uncertainty}
\label{app:scanners_boundaries}
SRC012 is derived from SkillFortifyBench. Under the final MEDIUM+ gate, excluding all SRC012-provenance units from the full benchmark changes SkillFortify Macro-F1 from 0.516 to 0.505. More importantly, excluding SRC012 from Source-Disjoint leaves SRC009+SRC011 and changes Macro-F1 from 0.349 to 0.254, malicious recall from 25.3\% to 16.7\%, and benign FPR from 49.9\% to 59.8\% (Table~\ref{tab:app_skillfortify_overlap}). This is a source-overlap sensitivity analysis, not evidence of training-set memorization.

\input{tables/appendix_skillfortify_overlap}

Exactly three technical abstentions occur across the 29,220 scanner--item jobs (Table~\ref{tab:app_scanner_abstains}). Two are Cisco failures---one roughly 21\,MB input for which no result is produced and one invalid-UTF-8 input---and SkillFortify reports no Skill for that same invalid-UTF-8 unit. All three occur in the Malicious-Structural-Disjoint test and none occurs in Source-Disjoint, so the main held-out-source comparison is unaffected by technical abstention.

\input{tables/appendix_scanner_technical_abstains}

Bootstrap intervals use 10,000 item-level resamples with seed 42 and quantify sampling uncertainty for deterministic scanner outputs; training-seed variance is outside this calculation. Table~\ref{tab:app_scanner_bootstrap} reports the Source-Disjoint intervals used to bound the paper-facing scanner comparison.

\input{tables/appendix_scanner_bootstrap_ci}

Cisco is additionally run in a documented local compatibility environment: scanner commit \texttt{48f59347a54b93606dd1e31c41989ebfd0fcc84d} declares \texttt{yara-x>=1.10}, whereas the frozen local environment contains \texttt{yara-x==1.4.0} installed with \texttt{--no-deps}. Scanner source/configuration is not modified and the configured static, bytecode, pipeline, and behavioral analyzers execute, but this dependency deviation limits exact replication of a dependency-clean upstream environment. Claims therefore apply to this frozen local compatibility configuration and do not generalize to every possible Cisco deployment.

\subsection{Threat-stratified malicious recall}
\label{app:scanners_attack}
\begin{figure}[H]
  \centering
  \includegraphics[width=\textwidth]{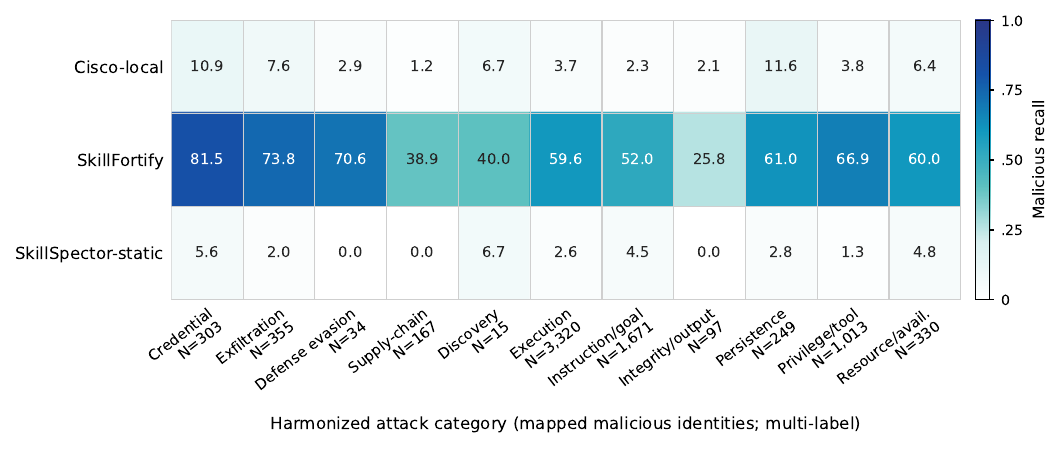}
  \caption{Malicious recall by harmonized attack category for the three frozen primary scanner configurations. Cell annotations are recall percentages; category sample sizes $N$ are shown on the x-axis. The analysis is conditional on the 4,983 malicious identities with supported attack mappings, and categories are multi-label. Differences are descriptive and may reflect source composition as well as scanner behavior.}
  \label{fig:scanner_attack_recall}
\end{figure}

Figure~\ref{fig:scanner_attack_recall} joins final primary scanner verdicts to the already-frozen harmonized attack mapping. The denominator is the same 4,983 attack-mapped malicious identities characterized in Appendix~\ref{app:threat}; categories are multi-label, so category sample sizes are not additive. SkillFortify has substantially higher recall than the other two primary scanner configurations across the mapped categories, including 81.5\% for Credential Access, 73.8\% for Data Exfiltration/Disclosure, 70.6\% for Defense Evasion/Obfuscation, 59.6\% for Execution/Code Delivery, and 66.9\% for Privilege/Tool/Authority Abuse. Its weakest mapped category is Integrity/Output Manipulation at 25.8\%. Cisco and SkillSpector remain low-recall across all categories at their frozen primary gates.

The stratified view should not be interpreted as a source-adjusted causal comparison. Attack categories are unevenly distributed across source datasets (Figure~\ref{fig:threat_landscape}b), and some categories have small $N$ (e.g., Discovery/Reconnaissance $N=15$ and Defense Evasion/Obfuscation $N=34$). We therefore use this analysis to describe coverage gaps, not to rank scanners by attack family independently of source composition.

\subsection{Comparability boundary}
The scanner study evaluates binary malicious-versus-benign detection on a common static artifact with fixed local configurations. It is not an adversarial-evasion/attack-success-rate experiment, and \textbf{SkillSpector-static} specifically uses its no-LLM mode. Results should therefore not be treated as numerically interchangeable with evaluations that use full packages, runtime behavior, cloud services, or LLM-backed scanner modes. The main conclusion is limited to the operational trade-off observed here: lower benign FPR can coincide with very low malicious recall, while a more sensitive scanner can incur substantial false positives.

%% file: tables/appendix_scanner_full_metrics.tex
\begin{table}[H]
\centering
\caption{Off-the-shelf scanner results on the full 9,740-unit benchmark. Primary metrics exclude technical abstentions and report coverage separately.}
\label{tab:app_scanner_full}
\small
\setlength{\tabcolsep}{4.2pt}
\begin{tabular}{lrrrrrr}
\toprule
Scanner & Coverage & Mal. recall & Benign FPR & Macro-F1 & FP & FN \\
\midrule
Cisco-local-behavioral & .9998 & .064 & .008 & .254 & 17 & 7,021 \\
SkillFortify-offline & .9999 & .566 & .451 & .516 & 1,007 & 3,257 \\
SkillSpector-static & 1.0000 & .033 & .002 & .222 & 4 & 7,258 \\
\bottomrule
\end{tabular}
\end{table}

%% file: tables/appendix_scanner_protocol_metrics.tex
\begin{table}[H]
\centering
\caption{Fixed scanner configurations on the four frozen protocol test sets. Scanners are not trained on protocol train splits; coverage is reported because technical abstentions remain outside primary binary metrics.}
\label{tab:app_scanner_protocols}
\small
\setlength{\tabcolsep}{4.0pt}
\begin{tabular}{llrrrr}
\toprule
Scanner & Protocol & Coverage & Mal. recall & Benign FPR & Macro-F1 \\
\midrule
Cisco-local & Random & 1.0000 & 0.065 & 0.004 & 0.254 \\
 & Source-Bal. & 1.0000 & 0.065 & 0.004 & 0.255 \\
 & M-Struct & 0.9990 & 0.082 & 0.004 & 0.272 \\
 & Source & 1.0000 & 0.025 & 0.011 & 0.308 \\
\addlinespace[1.5pt]
SkillFortify & Random & 1.0000 & 0.553 & 0.421 & 0.518 \\
 & Source-Bal. & 1.0000 & 0.569 & 0.490 & 0.504 \\
 & M-Struct & 0.9995 & 0.580 & 0.444 & 0.526 \\
 & Source & 1.0000 & 0.253 & 0.499 & 0.349 \\
\addlinespace[1.5pt]
SkillSpector-static & Random & 1.0000 & 0.033 & 0.002 & 0.222 \\
 & Source-Bal. & 1.0000 & 0.028 & 0.002 & 0.217 \\
 & M-Struct & 1.0000 & 0.047 & 0.000 & 0.237 \\
 & Source & 1.0000 & 0.000 & 0.006 & 0.281 \\
\bottomrule
\end{tabular}
\end{table}

%% file: tables/appendix_scanner_thresholds.tex
\begin{table}[H]
\centering
\caption{Pre-specified scanner operating-point sensitivity on the full benchmark. Primary rows are the paper-facing configurations.}
\label{tab:app_scanner_thresholds}
\small
\setlength{\tabcolsep}{4.5pt}
\begin{tabular}{llrrr}
\toprule
Scanner & Gate & Mal. recall & Benign FPR & Macro-F1 \\
\midrule
Cisco & HIGH+ / native (primary) & .064 & .008 & .254 \\
Cisco & MEDIUM+ & .127 & .025 & .310 \\
SkillFortify & MEDIUM+ (primary) & .566 & .451 & .516 \\
SkillFortify & native \texttt{is\_safe} & .576 & .474 & .514 \\
SkillFortify & HIGH+ & .565 & .450 & .515 \\
SkillSpector & block gate (primary) & .033 & .002 & .222 \\
SkillSpector & CAUTION+ & .151 & .042 & .329 \\
\bottomrule
\end{tabular}
\end{table}

%% file: tables/appendix_skillfortify_overlap.tex
\begin{table}[H]
\centering
\caption{SkillFortify source-overlap sensitivity under the fixed MEDIUM+ primary gate.}
\label{tab:app_skillfortify_overlap}
\small
\setlength{\tabcolsep}{5.0pt}
\begin{tabular}{lrrrr}
\toprule
Population & $N$ & Mal. recall & Benign FPR & Macro-F1 \\
\midrule
Full benchmark & 9,740 & .566 & .451 & .516 \\
Full excluding SRC012 provenance & 9,560 & .561 & .470 & .505 \\
Source-Disjoint & 1,384 & .253 & .499 & .349 \\
Source-Disjoint excluding SRC012 & 1,204 & .167 & .598 & .254 \\
\bottomrule
\end{tabular}
\end{table}

%% file: tables/appendix_scanner_technical_abstains.tex
\begin{table}[H]
\centering
\caption{The three technical abstentions among 29,220 scanner--item jobs. None occurs in Source-Disjoint; all appear in the Malicious-Structural-Disjoint test.}
\label{tab:app_scanner_abstains}
\small
\setlength{\tabcolsep}{3.4pt}
\begin{tabular}{lllll}
\toprule
Benchmark ID & Scanner & Label & Source & Technical reason \\
\midrule
\texttt{ASB04\_002434} & Cisco-local & Mal. & SRC001 & no result (21 MB input) \\
\texttt{ASB04\_007741} & Cisco-local & Ben. & SRC010 & invalid UTF-8 \\
\texttt{ASB04\_007741} & SkillFortify & Ben. & SRC010 & no Skill parsed \\
\bottomrule
\end{tabular}
\end{table}

%% file: tables/appendix_scanner_bootstrap_ci.tex
\begin{table}[H]
\centering
\caption{Bootstrap 95\% confidence intervals on Source-Disjoint (10,000 item-level resamples; seed 42). These intervals quantify sampling uncertainty for deterministic scanner outputs, not seed variance.}
\label{tab:app_scanner_bootstrap}
\small
\setlength{\tabcolsep}{4.2pt}
\begin{tabular}{lccc}
\toprule
Scanner & Macro-F1 & Mal. recall & Benign FPR \\
\midrule
Cisco-local & 0.308 [0.290, 0.325] & 0.025 [0.015, 0.036] & 0.011 [0.004, 0.021] \\
SkillFortify & 0.349 [0.324, 0.375] & 0.253 [0.224, 0.283] & 0.499 [0.457, 0.541] \\
SkillSpector-static & 0.281 [0.268, 0.295] & 0.000 [0.000, 0.000] & 0.006 [0.000, 0.013] \\
\bottomrule
\end{tabular}
\end{table}

%% file: appendix/appendix_examples.tex
\section{Representative Examples}
\label{app:examples}

This appendix makes the benchmark units and audit decisions concrete without reproducing operational malicious payloads. Every example is anchored to the frozen benchmark identifiers used elsewhere in the paper. For malicious artifacts we report source-side attack descriptors, provenance/evidence metadata, hashes or family identifiers when useful, and a non-operational abstraction of the content. Commands, endpoints, credentials, executable helper logic, and other payload-bearing details are omitted. Benign examples may be named more directly because they do not carry an attack payload.

\subsection{Representative benchmark units}
\label{app:examples_units}

Table~\ref{tab:app_representative_units} illustrates four distinct ways a primary benchmark unit can arise. The examples are not intended to define the entire taxonomy; they show why intent, provenance, evidence, artifact unit, and lineage are recorded separately. In particular, ``wild'' does not imply stronger evidence than ``constructed,'' and a backdoored package is not treated as interchangeable with an injected synthetic variant.

\input{tables/appendix_representative_units}

The malicious rows intentionally use safe abstractions and omit verbatim attack instructions. For example, the SRC011 record is retained as a wild malicious Skill because the frozen source identifies an exfiltration behavior under static evidence; the transport command and destination are not reproduced here. The SRC010 record is a constructed injected tool-hijack variant, while the SRC008 record is a constructed backdoored package. The benign SRC011 record demonstrates that the same canonical schema also preserves ordinary wild Skill artifacts as high-confidence reference negatives.

\subsection{A structural-reuse example}
\label{app:examples_structural}

Table~\ref{tab:app_structural_example} shows two distinct normalized malicious units assigned to the same frozen operational structural family. Their exact and normalized hashes differ, so this is not an exact/normalized duplicate case. The family contains five normalized identities and spans SRC001/SRC005/SRC010 after source-set propagation. In the independent blind positive-validation sample, the pair was judged high-confidence ``same structural template'': both retain the same YouTube-transcript summarization workflow and section ordering, while one variant adds a local prerequisite block. This is precisely the kind of scaffold reuse the structural partition is designed to capture.

\input{tables/appendix_structural_example}

The example also illustrates the interpretation boundary emphasized in Appendix~\ref{app:reuse_boundaries}: shared instructional structure does not establish a shared attacker, campaign, causal lineage, or attack mechanism. The family is an operational similarity control for evaluation, not an attack ontology.

\subsection{A cross-label conflict group}
\label{app:examples_conflict}

One conflict group from SRC006 (AgentTrap) makes the need for artifact-level conflict exclusion especially clear. Four source records share the same exact Skill hash and the same normalized hash, yet two source cases are canonicalized as benign test fixtures and two as malicious test fixtures. The shared base Skill is the same \texttt{xlsx} artifact; the case-level labels differ under the upstream runtime-fixture design. Table~\ref{tab:app_conflict_example} lists the four records without reproducing any executable fixture or payload.

\input{tables/appendix_conflict_example}

At the primary static-artifact level, automatically choosing one side would collapse source context into an unsupported ``correct'' artifact label. We therefore exclude the entire conflicting identity group from the primary benchmark, while retaining its provenance and evidence in the audit layer. This example should not be read as proof that either upstream label is erroneous: the conflict may arise because the source label is attached to surrounding runtime context while the benchmark unit contains only the static Skill text.

\subsection{A paired source-conditioned false-positive example}
\label{app:examples_source_shift}

The same frozen benign record can also behave differently under different training protocols. \texttt{ASB04\_007510} (\texttt{UASSB\_00181982}) is the SRC011 wild benign Skill \texttt{real-wix--wix-cli-backend-event}. It is a moderate/static, independently identified benign reference unit that appears in both frozen test manifests. The seed-42 word TF--IDF + linear-SVM model predicts the record correctly under Random but flips it to malicious under Source-Disjoint, even though the evaluated static Skill unit itself is unchanged. Table~\ref{tab:app_paired_fp} reports the frozen decision scores.

\input{tables/appendix_paired_fp}

This paired example provides a descriptive comparison under two protocols. The model parameters and training distributions differ between protocols, and source identity bundles content style, provenance, construction process, and label composition. The example therefore supports the paper's narrower claim of source-conditioned generalization sensitivity; it does not identify which source-associated factor caused the decision boundary to move.

%% file: tables/appendix_representative_units.tex
\begin{table}[H]
\centering
\caption{Representative primary benchmark units. Malicious content is summarized non-operationally; no payload command, endpoint, helper code, or executable instruction is reproduced.}
\label{tab:app_representative_units}
\scriptsize
\setlength{\tabcolsep}{4pt}
\renewcommand{\arraystretch}{1.10}
\begin{tabularx}{\textwidth}{@{}p{1.45cm}p{2.65cm}p{3.25cm}X@{}}
\toprule
Example & Frozen ID / source & Canonical metadata & Safe content abstraction \\
\midrule
Wild malicious & \texttt{ASB04\_002865}\\\texttt{UASSB\_}\\\texttt{00181999}\\SRC011 & wild; Skill text; moderate/static & Upstream Snyk exfiltration case. The source descriptor identifies exfiltration behavior; transport command and destination are omitted. \\
Injected malicious & \texttt{ASB04\_006504}\\\texttt{UASSB\_}\\\texttt{00175991}\\SRC010 & injected; Skill package; strong/constructed & SkillTrustBench case 00004 with a tool-hijack attack type; injected control-changing content is not reproduced. \\
Backdoored malicious & \texttt{ASB04\_006168}\\\texttt{UASSB\_}\\\texttt{00175105}\\SRC008 & backdoored; Skill package; strong/constructed; derived lineage & SkillTrojan EHR/SQL backdoor example. The hidden malicious delta is omitted; frozen base linkage is unresolved. \\
Benign reference & \texttt{ASB04\_009716}\\\texttt{UASSB\_}\\\texttt{00181530}\\SRC011 & wild; Skill text; moderate/static; independent lineage & ATR's official-fetch reference case, retained as a high-confidence Main-benign unit under the same canonical schema. \\
\bottomrule
\end{tabularx}
\end{table}

%% file: tables/appendix_structural_example.tex
\begin{table}[H]
\centering
\caption{Representative non-duplicate structural-reuse pair from a frozen operational structural family. The pair was included in the blinded positive-validation set and judged high-confidence same structural template.}
\label{tab:app_structural_example}
\scriptsize
\setlength{\tabcolsep}{4pt}
\renewcommand{\arraystretch}{1.10}
\begin{tabularx}{\textwidth}{@{}p{2.15cm}p{2.15cm}p{0.85cm}p{2.4cm}p{1.35cm}X@{}}
\toprule
Canonical / benchmark ID & Source-record shorthand & Source & Normalized hash prefix & Similarity to representative & Structural-review observation \\
\midrule
\texttt{UASSB\_}\\\texttt{00007858}\\\texttt{ASB04\_}\\\texttt{001347} & YouTube summarize (35o20) & SRC001 & \texttt{2c9012ef5e9f...} & 0.9967 & Shared transcript-summarization setup, extraction, metadata, summarization, language, and option sections. \\
\texttt{UASSB\_}\\\texttt{00007934}\\\texttt{ASB04\_}\\\texttt{003626} & yt-summarize & SRC001 & \texttt{7983b6100b3a...} & 0.9109 & Same ordered workflow; the blind reviewer identified a local prerequisite addition as the main structural difference. \\
\bottomrule
\end{tabularx}
\vspace{2pt}
\begin{minipage}{0.98\textwidth}
\scriptsize\emph{Family context.} The current family contains five normalized identities associated with SRC001/SRC005/SRC010. The historical raw-record context contains repeated released artifacts; these counts describe structural reuse, not independent attacks or campaigns.
\end{minipage}
\end{table}

%% file: tables/appendix_conflict_example.tex
\begin{table}[H]
\centering
\caption{Representative SRC006 cross-label identity conflict. All four records share exact Skill hash prefix \texttt{dd316db7785a...} and normalized hash prefix \texttt{6afac6531997...}; the conflicting group is excluded from the primary benchmark.}
\label{tab:app_conflict_example}
\scriptsize
\setlength{\tabcolsep}{4pt}
\renewcommand{\arraystretch}{1.10}
\begin{tabularx}{\textwidth}{@{}p{2.45cm}p{0.85cm}p{1.3cm}p{1.5cm}p{2.35cm}X@{}}
\toprule
Canonical ID & Record & Intent & Evidence & Upstream case descriptor & Shared artifact context \\
\midrule
\texttt{UASSB\_00107519} & 107 & benign & strong / runtime & DIM7: output tampering & \multirow{4}{=}{Derived from the same upstream \texttt{xlsx} Skill identity; artifact text is identical at the frozen exact-hash level.} \\
\texttt{UASSB\_00107520} & 108 & benign & strong / runtime & DIM8: unauthorized disclosure & \\
\texttt{UASSB\_00107538} & 20 & malicious & strong / runtime & DIM7: output tampering & \\
\texttt{UASSB\_00107550} & 22 & malicious & strong / runtime & DIM8: unauthorized disclosure & \\
\bottomrule
\end{tabularx}
\end{table}

%% file: tables/appendix_paired_fp.tex
\begin{table}[H]
\centering
\caption{Paired prediction for the same benign Skill under two frozen test protocols (seed-42 word TF--IDF + linear SVM). Positive decision scores predict malicious.}
\label{tab:app_paired_fp}
\scriptsize
\setlength{\tabcolsep}{5pt}
\renewcommand{\arraystretch}{1.10}
\begin{tabularx}{\textwidth}{@{}p{3.0cm}p{2.2cm}p{1.8cm}p{1.9cm}X@{}}
\toprule
Protocol & True label & Prediction & Decision score & Interpretation \\
\midrule
Random & benign & benign & $-0.360$ & Correct classification for \texttt{ASB04\_007510}. \\
Source-Disjoint & benign & malicious & $+0.509$ & False positive on the same unchanged static Skill unit under source-held-out training. \\
\bottomrule
\end{tabularx}
\vspace{2pt}
\begin{minipage}{0.98\textwidth}
\scriptsize\emph{Record metadata.} \texttt{UASSB\_00181982}, SRC011, \texttt{real-wix--wix-cli-backend-event}, \texttt{skill\_md}, wild provenance, moderate/static benign evidence, independent lineage. The exact seed-42 decision scores are $-0.360467$ (Random) and $+0.509328$ (Source-Disjoint).
\end{minipage}
\end{table}

%% file: appendix/appendix_release.tex
\section{Benchmark Release, Safety, and Governance}
\label{app:release}

\benchname is a security benchmark whose source registry contains both benign and malicious Agent Skill artifacts. Reproducibility therefore cannot be reduced to publishing every acquired byte in one archive. The release contract separates (i) the information needed to reproduce benchmark membership and evaluation, (ii) the public static-text representation of each benchmark identity, and (iii) source/package artifacts whose redistribution remains governed by recorded upstream terms. This appendix documents that contract and the maintenance rules for the frozen paper snapshot.

\subsection{Release package and reproducibility layers}
\label{app:release_layers}

The public benchmark release is organized into complementary layers for identities, text, metadata, and reproducibility artifacts. Table~\ref{tab:app_release_layers} summarizes the release-facing contract. All 9,740 benchmark identities have a publicly readable representation. Exact frozen static Skill text is released for 9,735 identities (7,500 malicious and 2,235 benign); the remaining five malicious identities have sanitized public representations because their exact originals contain sensitive credential material. Source/package redistribution beyond these per-identity text representations remains subject to recorded upstream terms and benchmark-side safety constraints.

\input{tables/appendix_release_layers}

The central design principle is that \emph{benchmark reproducibility is not identical to full-package republication}. Stable canonical IDs, pinned source revisions, exact and normalized hashes, source labels, provenance/evidence fields, conflict decisions, structural-family assignments, split manifests, and checksums make the paper snapshot auditable while the public text layer makes every primary identity inspectable. The 186 accepted identities whose text came from author-provided historical research snapshots (153 from SRC002 and 33 from SRC004) now include their exact frozen static Skill text in the public release; their historical provenance remains recorded. Full upstream packages or auxiliary source content may still require source-specific reacquisition under the recorded terms.

\subsection{Static-only handling and harmful-content minimization}
\label{app:release_safety}

All acquired Skill contents are treated as untrusted data. Benchmark construction, deduplication, structural comparison, text-feature extraction, sanitization, and the reported detector experiments operate on static text/metadata only. The pipeline does not execute Skill code, helper scripts, installers, embedded shell commands, URLs, network requests, evaluation harnesses, or payloads. This non-execution rule applies regardless of whether an upstream source describes an artifact as benign, malicious, vulnerable, synthetic, or a test fixture.

The same principle constrains release documentation. Reproducibility tables describe artifact identity, provenance, evidence, labels, transformations, and evaluation outcomes without requiring execution of operational payloads. For five credential-bearing malicious records, the released text is sanitized to remove sensitive credential material while preserving a readable Skill representation; these five texts are not bit-for-bit copies of the frozen experimental inputs. This release-safety exception is distinct from the deterministic scaffold-sanitization control in Appendix~\ref{app:robustness_sanitizer}; that control is an analysis condition with no effect on benchmark release identity and provides no guarantee that harmful semantics have been removed.

\subsection{Licensing and redistribution policy}
\label{app:release_licensing}

Each source retains its own upstream license and redistribution constraints. Appendix~\ref{app:sources_revisions} records the frozen revision and benchmark-side policy for all 13 sources. The benchmark does not infer redistribution permission from the fact that a repository is publicly accessible, and it does not use a permissive benchmark-level license to supersede a more restrictive source license.

For source/package material beyond the public per-identity text layer, we apply the following precedence rule: the released representation is the most informative form that is both consistent with the recorded upstream terms and compatible with the benchmark's safety policy. Practical cases include package redistribution where allowed, metadata/hash-only representation, or adapter-based reconstruction from the upstream source. Any later discovery of changed or clarified upstream terms is handled as a release-governance event while preserving the frozen scientific snapshot.

\subsection{Intended and out-of-scope uses}
\label{app:release_uses}

\textbf{Intended uses.} \benchname is designed for defensive and measurement-oriented research: static malicious Agent Skill detection, source-conditioned generalization evaluation, dataset provenance auditing, duplicate/reuse analysis, label-consistency auditing, benign false-positive analysis, threat-stratified analysis, and reproducibility studies over the frozen source registry. The benchmark can also support development of new defensive models provided that comparisons report the benchmark version, split protocol, detector inputs, and class-aware metrics.

\textbf{Out-of-scope uses.} The benchmark is not intended as an execution suite for malicious Skills, a payload-generation or payload-improvement resource, an actor/campaign attribution dataset, or a prevalence estimate for the wider Agent Skill ecosystem. Operational structural families are not attack-mechanism or campaign labels (Appendix~\ref{app:reuse_boundaries}), and Source-Disjoint is not a causal estimate of universal future-source performance (Appendix~\ref{app:evaluation_boundaries}). Likewise, performance of the static text baselines should not be interpreted as an upper bound on runtime, code-aware, behavioral, or multimodal security systems.

\subsection{Versioning, corrections, and maintenance}
\label{app:release_governance}

The benchmark state used in this paper is frozen for reproducibility. Future maintenance must preserve that snapshot, including its members, labels, hashes, structural-family assignments, and split manifests. Table~\ref{tab:app_governance_changes} specifies how common maintenance events should be handled.

\input{tables/appendix_governance_changes}

Every benchmark result should therefore identify at least: the benchmark release, the evaluation protocol, the detector input representation, the model/configuration, and the reported metric set. New source additions or revised labels can be scientifically useful, but they define a new benchmark release and should be evaluated separately from the frozen paper snapshot. This prevents a continuously updated corpus from silently changing the meaning of previously reported numbers.

\clearpage
\subsection{Compact benchmark card}
\label{app:release_benchmark_card}
Table~\ref{tab:app_benchmark_card} summarizes the frozen scientific and release-facing contract in one place. It is intentionally a compact index; the source-, construction-, threat-, protocol-, and scanner-specific appendices above provide the full details.

\input{tables/appendix_benchmark_card}

%% file: tables/appendix_release_layers.tex
\begin{table}[H]
\centering
\caption{Release layers for the frozen paper snapshot. Every primary identity has a public readable representation; source/package redistribution beyond that text layer remains source-specific as recorded in Appendix~\ref{app:sources_revisions}.}
\label{tab:app_release_layers}
\scriptsize
\setlength{\tabcolsep}{3.2pt}
\renewcommand{\arraystretch}{1.10}
\begin{tabularx}{\textwidth}{@{}>{\raggedright\arraybackslash}p{2.30cm}>{\raggedright\arraybackslash}p{5.00cm}>{\raggedright\arraybackslash}p{2.55cm}>{\raggedright\arraybackslash}X@{}}
\toprule
Layer & Contents & Release form & Reproducibility role \\
\midrule
Canonical registry & Canonical/source IDs, frozen revision, original/canonical label, provenance, evidence, label strength, artifact unit, lineage fields, license metadata & Public metadata & Reconstruct benchmark scope and source semantics \\
Identity and audit & Raw/exact identity hashes, normalized hashes, conflict flags, structural-family IDs, benchmark membership & Public metadata/hash & Verify identity, deduplication, reuse, and exclusions without executing artifacts \\
Primary static Skill text & Exact frozen text for 9,735 identities; sanitized readable representations for five credential-bearing malicious identities & Public text & Inspect all 9,740 primary identities while preserving the exact experimental text for 9,735 \\
Evaluation & Master unit IDs, split assignments, leakage checks, sanitizer rules, detector configurations, metric/result tables & Public non-payload artifacts & Reproduce paper protocols and reported analyses \\
Redistributable Skill artifacts & Static Skill text/packages whose recorded source terms and benchmark-side policy permit republication & Source-specific & Enable direct local detector input where redistribution is allowed \\
Restricted-source reconstruction & Pinned source revision, source record identifiers, adapter/reacquisition instructions, and checksums for artifacts not republished & Adapter or metadata/hash only & Let users reacquire under upstream terms and verify local correspondence \\
\bottomrule
\end{tabularx}
\end{table}

%% file: tables/appendix_governance_changes.tex
\begin{table}[H]
\centering
\caption{Maintenance and versioning contract. The paper snapshot remains immutable; scientifically material data changes create a new benchmark release while historical results remain tied to the frozen snapshot.}
\label{tab:app_governance_changes}
\scriptsize
\setlength{\tabcolsep}{3.4pt}
\renewcommand{\arraystretch}{1.10}
\begin{tabularx}{\textwidth}{@{}>{\raggedright\arraybackslash}p{3.10cm}>{\raggedright\arraybackslash}p{4.35cm}>{\raggedright\arraybackslash}X@{}}
\toprule
Maintenance event & Required action & Effect on frozen paper snapshot \\
\midrule
Documentation/citation clarification & Record in changelog; do not alter data bytes, labels, hashes, or splits & None; paper numbers remain directly comparable \\
License/redistribution clarification & Update release representation and attribution; preserve the recorded historical policy in the archived snapshot & May change what is distributed, not the historical benchmark membership \\
Label, content, or conflict correction & Issue a new benchmark release; regenerate affected canonical metadata, hashes, conflicts, splits, and derived metrics as needed & The paper snapshot remains available and unchanged \\
Source addition/removal or newer upstream revision & Create a new release with explicit source/revision delta and fresh deduplication/conflict auditing & Not backfilled into the paper snapshot \\
Split/protocol redesign & Version the protocol and keep the original assignments available & Prior results retain their original protocol meaning \\
New detector or leaderboard entry & Record model, inputs, configuration, release, protocol, and class-aware metrics & Does not require a benchmark-data version change \\
\bottomrule
\end{tabularx}
\end{table}

%% file: tables/appendix_benchmark_card.tex
\begin{table}[H]
\centering
\caption{Compact benchmark card for the frozen \benchname paper snapshot. Counts refer to the canonicalized static benchmark unless a subset is stated explicitly.}
\label{tab:app_benchmark_card}
\scriptsize
\setlength{\tabcolsep}{3.5pt}
\renewcommand{\arraystretch}{1.12}
\begin{tabularx}{\textwidth}{@{}>{\raggedright\arraybackslash}p{3.30cm}>{\raggedright\arraybackslash}X@{}}
\toprule
Field & Frozen paper-snapshot value / contract \\
\midrule
Source registry & 13 public sources; 182,699 canonical registry records; 11 sources contribute at least one Core malicious Skill artifact. \\
Core malicious construction & 8,414 raw artifacts $\rightarrow$ 7,562 exact-unique $\rightarrow$ 7,539 normalized-unique malicious identities. \\
Structural reuse control & 4,588 operational structural families at the frozen 0.68 threshold; family IDs support structural-disjoint evaluation and reuse auditing, with no campaign-attribution semantics. \\
Cross-label conflicts & 34 normalized malicious identities conflict with benign identities and are excluded from the primary benchmark under the frozen conflict policy. \\
Primary detection benchmark & 9,740 normalized-unique Skills: 7,505 malicious + 2,235 Main-benign. \\
Threat-characterization coverage & 4,983 malicious identities have harmonized multi-label attack mappings; 2,128 have conservative derived-impact mappings; 1,888 have both attack and impact mappings. \\
Frozen evaluation protocols & Random (6,818/974/1,948 train/dev/test); Source-Balanced Random (6,817/973/1,950); Malicious-Structural-Disjoint (6,818/974/1,948); Source-Disjoint (7,513/835/1,384), with 8 cross-boundary exclusions. \\
Detector suite & Three protocol-trained text baselines (word LR, word SVM, char SVM) plus three fixed off-the-shelf scanners (Cisco-local-behavioral, SkillFortify-offline, SkillSpector-static). \\
Artifact handling & Static analysis only: benchmark construction/evaluation never executes Skill instructions, helper code, installers, URLs, network actions, or payloads. \\
Public text release & All 9,740 identities have readable public representations; 9,735 provide exact frozen static Skill text (7,500 malicious + 2,235 benign), while five credential-bearing malicious identities provide sanitized representations only. \\
Release principle & Canonical metadata, hashes, split manifests, audit artifacts, and reproducibility material are public; full source/package redistribution beyond the per-identity text layer remains source-license- and safety-policy-specific. \\
\bottomrule
\end{tabularx}
\end{table}